\documentclass[review, 3p, authoryear]{elsarticle}
\usepackage{url}
\usepackage{hyperref}
 \hypersetup{
    colorlinks=true,
    linkcolor=blue,
    filecolor=blue,      
    urlcolor=blue,
}

\usepackage{natbib}

\usepackage{graphics,graphicx,multirow,dsfont}
\usepackage{amsmath,amssymb,verbatim,epsfig, amsthm}
\usepackage{booktabs}
\usepackage{multirow}
\usepackage{color, colortbl}
\usepackage{bbold}
\usepackage{subfig}

\newtheorem{proposition}{Proposition}

\newtheorem{defin}{\bf Definition}

\def\R{\mathds{R}}
\def\X{\mathds{X}}
\def\e{\mathrm{e}}

\def\ga{\mbox{Gamma}}
\def\iga{\mbox{inv-Gamma}}

\def\no{\mbox{N}}

\def\E{\mathds{E}}

\def\d{\mathrm{d}}
\def\rest{\mbox{rest}}

\def\bg{{\bf g}}

\def\bx{{\bf x}}

\def\bX{{\bf X}}
\def\bY{{\bf Y}}

\def\bzero{{\bf 0}}
\def\mt{{\tilde \mu}}
\def\simind{\stackrel{\mbox{ind}}{\sim}}
\def\simiid{\stackrel{\mbox{iid}}{\sim}}

\newcommand{\bgamma}{\boldsymbol{\gamma}}
\newcommand{\bdelta}{\boldsymbol{\delta}}
\newcommand{\btheta}{\boldsymbol{\theta}}

\newcommand{\PP}{\mathcal{P}}

\DeclareMathOperator{\argmin}{\arg\min\,}
\allowdisplaybreaks

\makeatletter
\def\ps@pprintTitle{%
  \let\@oddhead\@empty
  \let\@evenhead\@empty
  \def\@oddfoot{\reset@font\hfil\thepage\hfil}
  \let\@evenfoot\@oddfoot
}
\makeatother

\begin{document}

\title{Optimal stratification of survival data via Bayesian nonparametric mixtures} 
\journal{Econometrics and Statistics}

\author{Riccardo Corradin\corref{cor1}}
\ead{riccardo.corradin@unimib.it}
\address{Department of Economics, Management and Statistics, University of Milano-Bicocca\\  Piazza dell'Ateneo Nuovo, 1, 20126, Milano, Italy}
\author{Luis Enrique Nieto-Barajas}
\address{Department of Statistics, ITAM\\ 
Río Hondo No. 1, Col. Progreso Tizap\'an,
01080 Alc. \'Alvaro Obreg\'on, Ciudad de México, México}
\author{Bernardo Nipoti}
\address{Department of Economics, Management and Statistics, University of Milano-Bicocca\\[4pt] Piazza dell'Ateneo Nuovo, 1, 20126, Milano, Italy}
\cortext[cor1]{Corresponding author}

\begin{abstract}
{The stratified proportional hazards model represents a simple solution to account for heterogeneity within the data while keeping the multiplicative effect on the hazard function. Strata are typically defined a priori by resorting to the values taken by a categorical covariate. 
A general framework is proposed, which allows for the stratification of a generic accelerated life time model, including as a special case the Weibull proportional hazard model. The stratification is determined a posteriori by taking into account that strata might be characterized by different baseline survivals as well as different effects of the predictors. This is achieved by considering a Bayesian nonparametric mixture model and the posterior distribution it induces on the space of data partitions. The optimal stratification is then identified by means of the variation of information criterion and, in turn, stratum-specific inference is carried out. The performance of the proposed method and its robustness to the presence of right-censored observations are investigated by means of an extensive simulation study. A further illustration is provided by the analysis of a data set extracted from the University of Massachusetts AIDS Research Unit IMPACT Study.}
\end{abstract}


\maketitle

\section{Introduction}
\label{sec:intro}

In the study of lifetime data with covariates, the proportional hazards \citep{cox:72} and the accelerated life  \cite[e.g.][]{cox&oakes:84} are arguably the most popular models. To state these models in notation, let $(T_i,\bX_i)$ be a set of observable random variables for individual $i$ such that $T_i$ is a non-negative (lifetime) variable and $\bX_i$ is a vector of explanatory variables. The proportional hazards model assumes that the effect of the covariates is multiplicative on the hazard rate, i.e. 
\begin{equation}
\nonumber
h(t_i\mid\bx_i)=\e^{\btheta'\bx_i}h_0(t_i),
\end{equation}
where $h(t_i\mid\bx_i)$ is the hazard rate function of individual $i$, $h_0(t)$ is a baseline hazard rate function common to all individuals, and $\btheta$ is a vector of regression coefficients. The accelerated life model, on the other hand, assumes that the effect of the covariates is multiplicative in the lifetime, which in terms of the hazard rate can be written as 
\begin{equation}
\label{eq:al0}
h(t_i\mid\bx_i)=\e^{\btheta'\bx_i}h_0\left(\e^{\btheta'\bx_i}t_i\right).
\end{equation}
In the latter, the covariates have the effect of accelerating the survival time $T_i$ when $\btheta'\bx_i>0$, whereas $\btheta'\bx_i<0$ induces a delay of the same. 

It is common practice to generalize the proportional hazards model to include heterogeneity in the baseline hazard by considering a different hazard function for each of several predefined groups or strata, while keeping the multiplicative effect on the hazard function. Let $Z_i$ be a categorical or discrete variable taking values in a set $\{1,\ldots,J\}$, then the stratified proportional hazards model \citep[e.g.][]{kalbfleish&prentice:02,lawless:03} for individual $i$ is defined as
\begin{equation}
\label{eq:sph}
h(t_i\mid\bx_i,z_i)=\e^{\btheta_{z_i}'\bx_i}h_{z_i}(t_i),
\end{equation}
where $h_j(t)$ is the baseline hazard for stratum $j$, with $j=1,\ldots,J$, and $\btheta_j$ is the vector of regression coefficients, either assumed to be stratum-specific, which implies an interaction between $\bx_i$ and $z_i$, or constant, i.e. $\btheta_j=\btheta$, which implies a common effect across strata. 

Although not much explored in the statistical literature, there is no reason against defining a stratified version of the accelerated life model. 
Such an extension can be written as 
\begin{equation}
\label{eq:sal}
h(t_i\mid\bx_i,z_i)=\e^{\btheta_{z_i}'\bx_i}h_{z_i}\left(\e^{\btheta_{z_i}'\bx_i}t_i\right).
\end{equation}
Again, the regression coefficients $\btheta_j$ can be assumed either different or equal across strata.

The stratified regression models \eqref{eq:sph} and \eqref{eq:sal} define a stratification through a categorical or discrete variable but such definition of the strata, determined \emph{a priori}, might not actually be supported by the data. Additionally, a more convenient stratification might arise from the combination of several variables. Here we propose a way of defining, \emph{a posteriori}, an optimal stratification of the data, which, unlike standard procedures, clusters observations on the basis of the effect of the covariates and the observed survivals, rather than the value of the covariates themselves. We achieve this by means of a Bayesian nonparametric mixture model. The latter has the appealing property of considering the number of strata random and allows it to be inferred conditionally on the observation of the data. Additionally, available information on the number of strata can be incorporated into the model by tuning the prior distribution for the number of strata. Our contribution adds to an existing body of literature on nonparametric mixtures of accelerated life models for data analysis and clustering \citep{Arg09, Arg10, ArD14,Liv20}.

Bayesian nonparametric mixture models relying on an almost surely discrete measure, such as the Dirichlet process (DP) \citep{ferguson:73}, the Pitman--Yor process \citep{per:92}, normalized random measures \citep{reg:03} and the stick breaking processes \citep{Ish01}, among others \citep[see, e.g.,][]{hjort&al:10}, have proved useful in identifying clusters of individuals. We exploit this feature to produce a stratification based on all available covariates and, in turn, carry out inference for each stratum conditionally on the inferred groups.

The literature on regression methods via nonparametric mixtures is rich. \citet{DeI04} define dependent mixture models indexed by the values taken by a categorical covariate. The same idea has been largely exploited in other papers, with the regression approach extended as to account for continuous explicative variables as well \citep[see, e.g.,][]{Dun07}. Nonparametric mixtures have been tailored to model survival data by choosing suitable kernels, such as gamma \citep{Han06}, Weibull \citep{Kot06} and Burr \citep{Haj18}. Alternatively, nonparametric mixtures based on the use of mixing completely random measures have been deployed to model hazard functions \citep{Dyk81,nieto&walker:04}, possibly accounting for the presence of covariates \citep[see, e.g.,][]{nieto&walker:05,Nip18}. Nonparametric mixtures for accelerated life models in the presence of covariates have been considered in \citet{ArD14} and \citet{Liv20}, with the goal of producing cluster-specific posterior inference. The first reference --see also \citet{Arg09,Arg10} for allied work-- proposes the definition of Weibull accelerated life models with mixing normalized generalized gamma process and models the effect of predictors as common across clusters; the second one models survival data with a DP mixture with generic kernel admitting a convenient factorization into the product of a response and a covariate model.

We propose a general framework that allows for the stratification of a generic accelerated life time model, including as a special case the popular Weibull proportional hazards model. The mixture model we present is defined in great generality, by means of a generic kernel function and a mixing normalized random measure. The approach we propose accounts for cluster-specific covariate effect, thus making the optimal stratification of the data depend also on the different effect that predictors might have on individual responses. Generating a sample from the posterior distribution of the model on the space of partitions, by means of Markov chain Monte Carlo, represents the first step towards the identification of an optimal partition, which is then obtained by resorting to the variation of information criterion presented by \citet{Wad18}. Our model naturally accounts for the presence of possibly censored survival data: with an extensive simulation study we will shed light on the robustness of the methods we consider in producing an accurate stratification of the data in the presence of censoring. 

The remaining of the paper is organised as follows. A general mixture model, embedded within accelerated life or proportional hazard models, is presented in Section~\ref{sec:model} along with the list of kernels that will be considered when the model is actually implemented. Section~\ref{sec:post} introduces all is needed to carry out posterior inference on the stratification and on stratum-specific quantities of interest, including the complete specification of the prior model via the choice of a normalized inverse Gaussian process, the full conditional distributions for posterior sampling, and the variation of information criterion for identifying the optimal stratification. In Section~\ref{sec:simu} an extensive simulation study is presented to assess the ability of the proposed methods in detecting suitable stratifications of the data and to test their robustness to the presence of right-censored data. The proposed methodology is further illustrated in Section~\ref{sec:illustration} by means of the analysis of a data set extracted from the University of Massachusetts
AIDS Research Unit IMPACT Study. Concluding remarks are presented in Section~\ref{sec:discussion}, while further details on the model, the proof of Proposition~\ref{prop:posterior}, 
and additional results produced by the simulation study of Section \ref{sec:simu} and by the illustration of Section \ref{sec:illustration}, are available in the Appendix.

\section{Unified framework}
\label{sec:model}

The stratified proportional hazards and the stratified accelerated life models become fully parametric if $h_0(t)$ is given a parametric family, otherwise the model remains semiparametric. It is well known \citep[][p. 71]{cox&oakes:84} that when the baseline hazard is Weibull the proportional hazards and the accelerated life models coincide, provided the covariates are constant. Therefore, in order to keep a common framework for both models, we will consider the accelerated life model for several choices of $h_0(t)$ and recover the proportional hazards model when selecting $h_0$ to be Weibull. The resulting class of proportional hazards models with Weibull baseline hazard is very flexible and therefore a popular choice in applications \citep[e.g.][]{zhang:16,Val17}.

An alternative representation of the accelerated life model \eqref{eq:al0}, expressed in terms of random variables, is obtained by defining $T_i=T_0/\e^{\btheta' \bx_i}$, where the distributions of $T_i$ and $T_0$ are characterized respectively by hazard functions $h_i$ and $h_0$. 
It is worth noticing that, in order to achieve model identifiability, the vector of regression coefficients $\btheta$ must not include an intercept term.
A logarithmic transformation leads to $\log T_i=\mu-\btheta'\bx_i+\epsilon_0$, where $\mu=\E(\log T_0)$ and $\epsilon_0=\log T_0-\mu$ is a zero mean random variable. If we further define $Y_i=\log T_i$, 
and $\zeta Y_0=\epsilon_0$, with $Y_0$ being a random variable with standard distribution (zero mean and fixed variance) and $\zeta>0$, then the accelerated life model is written as a log-linear model of the form $Y_i=\mu-\btheta'\bx_i+\zeta Y_0$. Henceforth an asterisk will be used to denote probability functions referring to the logarithm of survival times, such as $Y_0$ and $Y_i$, in order to avoid confusion with probability functions referring to the survival times $T_0$ and $T_i$. We thus denote by $S^*_0(y)$ the survival function of $Y_0$, and observe that the survival function $S(t_i)$ of $T_i=\e^{Y_i}$ is given by $$S(t_i)=S^*_0\left(\frac{\log t_i-\mu+\btheta'\bx_i}{\zeta}\right).$$

While several choices can be made, we focus our attention on three specific functions $S_0^*(y)$, which are the most widely used parametric lifetime models belonging to the class of log-location-scale distributions \citep[see e.g.][p. 211]{lawless:03}. Along with the specification of $S^*(y)$, we display the corresponding density $f^*(y)$ and the distribution of $T_i$. 

\medskip
\noindent \textit{Kernel I: type-I minimum distribution for $Y_i$ (Weibull distribution for $T_i$).}\\
\noindent We let $Y_0\sim \text{Ev}_\text{I}(\tilde{\gamma} \sqrt{6}/\pi,\sqrt{6}/\pi)$, where $\text{Ev}_{\text{I}}$ denotes the type-I minimum distribution and $\tilde{\gamma}\approx 0.5772$ is the Euler-Mascheroni constant, that is $Y_0$ is a random variable with mean $0$ and variance $1$, with survival function given by
\begin{align*}
S_0^*(y)&=\exp\left\{-\e^{\frac{\pi}{\sqrt{6}}y-\tilde{\gamma}}\right\}.
\end{align*}
Then we have that 
\begin{align*}
Y_i\mid \mu, \btheta, \zeta, \bx_i  \sim&\; \text{Ev}_\text{I}\left(\mu-\btheta^\prime \bx_i+\zeta \frac{\tilde{\gamma}\sqrt{6}}{\pi},\zeta \frac{\sqrt{6}}{\pi}\right),\\
T_i\mid \mu, \btheta, \zeta, \bx_i  \sim&\; \text{Weibull}\left(\e^{\mu-\btheta^\prime \bx_i+\zeta\frac{\tilde{\gamma} \sqrt{6}}{\pi}},\frac{\pi}{\zeta\sqrt{6}}\right).
\end{align*}
The density of $Y_i$ is thus given by
\begin{equation}\label{eq:densEvi}
f^*(y\mid \mu, \boldsymbol{\theta},\zeta,\mathbf{x}_i)=\exp\left\{-\e^{\pi\frac{y-\mu+\boldsymbol{\theta}^\prime \mathbf{x}_i}{\zeta \sqrt{6}}-\tilde{\gamma}}+\pi\frac{y-\mu+\boldsymbol{\theta}^\prime \mathbf{x}_i}{\zeta\sqrt{6}}-\tilde{\gamma}\right\}\frac{\pi}{\zeta\sqrt{6}}.
\end{equation}

\medskip
\noindent\textit{Kernel II: logistic distribution for $Y_i$ (log-logistic distribution for $T_i$).}\\
\noindent We let $Y_0\sim \text{Logistic}(0,\sqrt{3}/\pi)$, that is $Y_0$ is a random variable with mean $0$ and variance $1$, with survival function given by
\begin{align*}
S_0^*(y)&=1-\left(1+\exp\left\{-\frac{\pi }{\sqrt{3}}\,y\right\}\right)^{-1}.
\end{align*}
Then we have that 
\begin{align*}
Y_i\mid \mu,\btheta,\zeta,\bx_i \sim&\; \text{Logistic}\left(\mu-\btheta^\prime \bx_i,\zeta \frac{\sqrt{3}}{\pi}\right)\\
T_i\mid \mu,\btheta,\zeta,\bx_i  \sim& \; \text{log-Logistic}\left(\e^{\mu-\btheta^\prime \bx_i},\frac{\pi}{\zeta \sqrt{3}}\right).
\end{align*}
That is the density of $Y_i$ is given by 
\begin{equation}\label{eq:densLog}
f^*(y\mid \mu,\btheta,\zeta,\bx_i)=\frac{\pi}{\zeta \sqrt{3}}\,\frac{\exp\left\{-\pi\frac{y-\mu+\btheta^\prime \bx_i}{\zeta \sqrt{3}}\right\}}{\left(1+\exp\left\{-\pi\frac{y-\mu+\btheta^\prime \bx_i}{\zeta \sqrt{3}}\right\}\right)^2}.
\end{equation}

\medskip
\noindent \textit{Kernel III: normal distribution for $Y_i$ (log-normal distribution for $T_i$).}\\
\noindent We let $Y_0\sim \text{N}(0,1)$, that is $Y_0$ is a standard normal random variable 
whose cumulative distribution function is denoted as
\begin{align*}
S_0^*(y)&=1-\Phi\left(y\right).
\end{align*}
Then we have that 
\begin{align*}
Y_i\mid \mu,\btheta,\zeta,\bx_i  \sim&\; \text{N}\left(\mu-\btheta^\prime \bx_i,\zeta^2\right)\\
T_i\mid \mu,\btheta,\zeta,\bx_i \sim&\; \text{log-N}\left(\mu-\btheta^\prime \bx_i,\zeta^2\right).
\end{align*}
That is the density of $Y_i$ is given by 
\begin{equation}\label{eq:densN}
f^*(y\mid \mu,\btheta,\zeta,\bx_i)=\frac{1}{\zeta}\varphi\left(\frac{y-\mu+\btheta^\prime \bx_i}{\zeta}\right).
\end{equation}

In all cases $\mu$ and $\zeta$ are the location and scale parameters of the baseline distribution of the accelerated life model, while $\btheta$ determines the effect of the predictors. In order to create a stratified version of the model, as displayed in \eqref{eq:sal}, we induce ties among individual parameters $(\mu,\btheta,\zeta)$ and interpret individuals sharing the same value of $(\mu,\btheta,\zeta)$ as belonging to the same stratum. 
This can be achieved by allowing each individual to have their specific values for the parameters $(\mu,\btheta,\zeta)$, say $(\mu_i,\btheta_i,\zeta_i)$, and by assuming these parameters are exchangeable from a nonparametric discrete distribution. As a result, we will be able to identify an optimal stratification by using information from the observed data and not a priori. It is worth stressing that the approach we propose, henceforth referred to as model M2, takes into account the effect of the covariates when defining the optimal stratification of the data, so that individuals in different strata might be characterized by different baseline distributions as well as different regression coefficients. As far as the regression coefficients are concerned, two alternative modelling strategies will be considered as well for the purpose of comparison. A first approach, referred to as model M0, assumes that the covariates have no effect on the responses, that is that $\btheta_i=\bzero$ for every $i=1,\ldots,n$. A second option, along similar lines to \citet{ArD14} and named here model M1, consists in assuming the effect of the covariates is common to all individuals across strata, that is $\btheta_i=\btheta$ for all $i=1,\ldots,n$, with independent priors for the coefficients.

In summary, the three modelling alternatives for the log-times-to-event $Y_i=\log T_i$ that we consider can be formalized as alternative specifications of a Bayesian nonparametric mixture model. Namely,
\begin{enumerate}
\item[M0.]
Null covariates effect model:
\begin{eqnarray}
\nonumber
Y_i\mid\mu_i,\zeta_i,\mathbf{x}_i & \simind & f^*(y_i\mid\mu_i,\btheta=\bzero,\zeta_i,\mathbf{x}_i),\quad i=1\ldots,n \\
\label{eq:bnpmm0}
\bgamma_i=(\mu_i,\zeta_i)\mid G & \simiid & G, \quad i=1,\ldots,n \\
\nonumber
G & \sim & \PP(G), 
\end{eqnarray}
\item[M1.]
Common $\btheta$ model:
\begin{eqnarray}
\nonumber
Y_i\mid\mu_i,\btheta,\zeta_i,\mathbf{x}_i & \simind & f^*(y_i\mid\mu_i,\btheta,\zeta_i,\mathbf{x}_i),\quad i=1\ldots,n \\
\nonumber
\btheta & \sim & \pi(\btheta), \\
\label{eq:bnpmm1}
\bgamma_i=(\mu_i,\zeta_i)\mid G & \simiid & G, \quad i=1,\ldots,n \\
\nonumber
G & \sim & \PP(G), 
\end{eqnarray}
\item[M2.] Individual $\btheta_i$ model:
\begin{eqnarray}
\nonumber
Y_i\mid\mu_i,\btheta_i,\zeta_i,\mathbf{x}_i & \simind & f^*(y_i\mid\mu_i,\btheta_i,\zeta_i,\mathbf{x}_i),\quad i=1\ldots,n \\
\label{eq:bnpmm2}
\bgamma_i=(\mu_i,\btheta_i,\zeta_i)\mid G & \simiid & G, \quad i=1,\ldots,n \\
\nonumber
G & \sim & \PP(G), 
\end{eqnarray}
\end{enumerate}
where $f^*$ is specified as \eqref{eq:densEvi}, \eqref{eq:densLog} or \eqref{eq:densN}, $\pi(\btheta)$ is a multivariate prior, $G$ is a random probability measure and $\PP$ is a nonparametric, almost surely discrete, prior for $G$. In this work, we focus on the class of homogeneous normalized random measures with independent increments (NRMI), although a similar framework can be devised for other nonparametric priors such as, for example, the Pitman--Yor process, as hinted at in \ref{sec:PY}. It is worth stressing that the parameters $\bgamma_i$ in \eqref{eq:bnpmm0}, \eqref{eq:bnpmm1} and \eqref{eq:bnpmm2} are specific to each individual $i$. However, since the prior $\PP$ is almost surely discrete, the vectors $\bgamma_i$, $i=1,\ldots,n$ will feature ties with positive probability thus producing a stratification in the data. Once posterior inference is carried out by analysing a sample generated from the posterior distribution of $\bgamma=(\bgamma_1,\ldots,\bgamma_n)$, an optimal stratification will be chosen by means of the variation of information criterion of \citet{Wad18}. The generated posterior sample can also be used to produce stratum-specific posterior inference. Alternatively, as done in this paper, an assumption of independence across strata can be formulated and, conditionally on the identified stratification, the same nonparametric mixture model used to determine the optimal stratification can be implemented independently for each stratum.

\section{Prior to posterior analysis}\label{sec:post}

Let $\bY=(Y_1,\ldots,Y_n)$ be a sample of size $n$ from model \eqref{eq:bnpmm0}, \eqref{eq:bnpmm1} or \eqref{eq:bnpmm2}. As common in survival analysis, we account for the fact that survival times $T_i$, and thus $Y_i$, might not be observed exactly, with only partial information available. We focus on right-censoring here since it is the most frequent in practical applications, although extending the framework we devise for left or interval censored observations is straightforward. We further assume that censoring times are independent of survival times. We use the censoring indicator $\delta_i$ to distinguish an exact observation ($\delta_i=1$) from a right-censored observation ($\delta_i=0$). Besides the variable of interest $Y_i$, we also observe a vector of covariates $\bX_i$. In summary, data consist of triplets $(Y_i,\delta_i,\bX_i)$, for $i=1,\ldots,n$, that is $(\bY,\bdelta,\bX)$ where, for the sake of simplicity, we introduced the notation $\bdelta=(\delta_1,\ldots,\delta_n)$ and $\bX=(\bX_i,\ldots,\bX_n)$.
In what follows we describe the nonparametric prior distribution $\PP$ of $G$ with the aim of investigating the marginal posterior distribution of $\bgamma$, and thus the distribution on the space of data partitions implied by $\bgamma$, given the observations $(\bY,\bdelta,\bX)$, where the marginalization is intended with respect to $G$.  The joint posterior distribution of $(\bgamma,\bY,\bdelta)$, conditionally on $\bX$, represents the starting point to derive the full conditional distributions needed to devise a Gibbs sampler for posterior inference. 

\subsection{Prior distribution}\label{sec:prior}
We denote by  $m$ the dimension of each $\bgamma_i$, whose components are the location parameter $\mu_i$ and the scale parameter $\zeta_i$ in models M0 and M1, and the location parameter $\mu_i$, the regression coefficients $\btheta_i$ and the scale parameter $\zeta_i$ in model M2. 
We suppose that $\PP(G)$ is the law of a homogeneous NRMI, that is we assume that the random probability measure $G$ is defined as $G(\cdot)=\mt(\cdot)/\mt(\X)$ where $\mt$ is a completely random measure defined on the space $\X=\R^{m-1}\times \R^+$, such that $0<\tilde\mu(\X)<\infty$ almost surely. Without loss of generality, we also assume that $\mt$ has no fixed points of discontinuity. The law of a homogeneous $\mt$, and thus the law of $G$, is characterized by a L\'evy intensity of the form
\begin{equation}
\label{eq:levy}
\nu(\d s,\d x)= \rho(s) \d s\, \alpha G_0(\d x),
\end{equation}
where $\rho$ is a non-negative function, 
$\alpha$ is a positive constant 
and $G_0$ is a diffuse probability measure on $\X$. 
An important feature of $\mt$ is its almost sure discreteness, which implies that it can be written as $\mt(\cdot)=\sum_{j\geq 1}J_j \delta_{\tilde{X}_j}(\cdot)$, with the homogeneity assumption \eqref{eq:levy} translating into independence of jumps $J_j$ and locations $\tilde{X}_j$. While the locations are independent and identically distributed from $G_0$, the distribution of the random jumps can be described in terms of $\rho$ \citep{Fer72}. As a result, also the random probability measure $G$ is almost surely discrete and can be represented as $G(\cdot)=\sum_{j\geq 1}\tilde{J}_j \delta_{\tilde{X}_j}(\cdot)$, where $\tilde{J}_j=J_j/\sum_{i\geq 1}J_i$. Moreover, since $\E(G)=G_0$, the probability measure $G_0$ can be interpreted as centering measure or prior guess of $G$. Finally, the Laplace functional transform of $\mt$ is characterized by $\E\left[\e^{-\lambda\mt(A)}\right]=\e^{-\alpha G_0(A)\psi(\lambda)}$, for any measurable $A\subseteq\X$ and any $\lambda>0$, with $\psi(\lambda):=\int_{\R_+} (1-\e^{-s\lambda})\rho(s)\d s$ known as the Laplace exponent of $\mt$. 
For a comprehensive introduction to completely random measures we refer to \citet{Dal08} and \citet{Lij10}, with the latter providing an overview of their use in Bayesian nonparametric statistics. 

\subsection{Posterior sampling}\label{sec:post_sampling}

We next focus on model M2 and derive explicit expressions for the full conditional distributions of the elements of $(\bgamma,\bY,\bdelta)$. The same expressions are easily adapted to the case of observations from models M0 and M1. 
Such full conditionals are needed to implement the Gibbs sampling algorithm to carry out posterior inference, and are easily obtained once the joint distribution of $(\bgamma,\bY,\bdelta)$, conditionally on $\bX$, is available. 
A convenient characterization of the latter is provided bt \citet{Jam09}, who also study the posterior distribution of $G$ \citep[see, also,][for a review and for a detailed account of possible computational approaches]{Fav13}. We display here an extension of the aforementioned result, accounting for the presence of possibly right-censored observations. Before stating the next proposition, we observe that, due to the almost sure discreteness of $\mt$, and thus of $G$, the vectors $\bgamma_i$ will display ties with positive probability, giving rise to $k\leq n$ distinct values $(\bgamma_1^*,\ldots,\bgamma_k^*)$, with frequencies $(n_1,\ldots,n_k)$ such that $\sum_{i=1}^k n_i=n$.

\begin{proposition}
\label{prop:posterior}
Let $(Y_i,\delta_i,\bX_i)$, $i=1,\ldots,n$, be a set of observable random variables from model M2. Let $G$ be a homogeneous NRMI with L\'evy intensity \eqref{eq:levy}. Then the joint conditional density of $(\bgamma,\bY,\bdelta,U)$ given $\bX$, with $U$ being a suitable auxiliary variable, is given by
\begin{align}
\nonumber
\frac{\alpha^k}{\Gamma(n)} u^{n-1}\e^{-\alpha\psi(u)}\prod_{j=1}^{k} \kappa_{n_j}(u) G_0(\d \bgamma_j^*) \prod_{i\in C_j} f^*(y_i\mid \bgamma_j^*,\bx_i)^{\delta_i} S^*(y_i\mid \bgamma_j^*,\bx_i)^{1-\delta_i},
\end{align}
where the sets $C_j=\left\{i\in\{1,\ldots,n\} : \bgamma_i =\bgamma_j^*\right\}$ and $\kappa_{n_j}(t):=\int_{0}^{\infty}\e^{-s t}s^{n_j}\rho(s) \d s$.
\end{proposition}

For the sake of completeness, the proof is reported in \ref{sec:proof1}.

\subsection{Full conditional distributions}\label{sec:fc_NRMI}

Starting from the joint distribution in Proposition~\ref{prop:posterior}, we assume that the parameter $\alpha$ is assigned a Gamma prior distribution with shape and rate parameters denoted by $q_0^{(\alpha)}$ and $q_1^{(\alpha)}$ respectively, and we write the full conditional distributions for the random elements $(\bgamma,\alpha,U)$. The latter represent the building block for the Gibbs sampling algorithm implemented for the analyses displayed in Sections~\ref{sec:simu} and~\ref{sec:illustration}, which consists of an adaptation to the case of possibly right-censored data of the nonconjugate marginalized sampler of \citet{Fav13} (see their Section 3.2).
\begin{enumerate}
\item[(a)] The full conditional of $\alpha$ is given by
$$\alpha\mid\rest\sim\ga(q_0^{(\alpha)}+k,q_1^{(\alpha)}+\psi(u)).$$
\item[(b)] The full conditional of $U$ is given, for any $u>0$, by 
$$f(u\mid \rest)\propto u^{n-1}\e^{-\alpha\,\psi(u)}\prod_{j=1}^k \kappa_{n_j}(u).$$
\item[(c)] The full conditional distribution for each vector $\bgamma_i$, $i=1,\ldots,n$, is given by 
\begin{equation}\label{eq:fc_gamma}
f(\bgamma_i \mid \rest)=p_{0,i} f_{0,i}(\bgamma_i) + \sum_{j=1}^{k^{(i)}}p_{j,i} \mathds{1}_{\bgamma_{j,i}^{*}}(\bgamma_i),
\end{equation}
where $k^{(i)}$ is the number of distinct values featured by the vector obtained by excluding $\bgamma_i$ from $\bgamma$, that is $\left(\bgamma_{1,i}^{*} ,\ldots,\bgamma_{k^{(i)},i}^{*}\right)$, with cardinalities $\left(n_{1,i},\ldots,n_{k^{(i)},i}\right)$, and the weights are given by
\begin{equation}\label{eq:new}
p_{0,i}\propto \alpha \kappa_{1}(u)\int_{\R^m}f^*(y_i\mid \bg,\bx_i)^{\delta_i} S^*(y_i\mid \bg,\bx_i)^{1-\delta_i} g_0(\bg)\d\bg,
\end{equation}
and for $j=1,\ldots,k^{(i)}$,
$$p_{j,i}\propto \frac{\kappa_{n_{j,i}+1}(u)}{\kappa_{n_{j,i}}(u)}f^*(y_i\mid \bgamma_{j,i}^*,\bx_i)^{\delta_i}S^*(y_i\mid \bgamma_{j,i}^*,\bx_i)^{1-\delta_i}.$$
Finally 
\begin{equation*}f_{0,i}(\bgamma)=\frac{f^*(y_i\mid \bgamma,\bx_i)^{\delta_i} S^*(y_i\mid \bgamma,\bx_i)^{1-\delta_i} g_0(\bgamma)}{\int_{\R^m}f^*(y_i\mid \bg,\bx_i)^{\delta_i} S^*(y_i\mid \bg,\bx_i)^{1-\delta_i} g_0(\bg)\d\bg},\end{equation*}

where $g_0$ is the density function associated with the probability measure $G_0$.

\end{enumerate}

Given that any kernel $f^*$ among \eqref{eq:densEvi}, \eqref{eq:densLog} and \eqref{eq:densN}, might be used, we implement a sampling scheme for the posterior distribution which does not rely on the conjugacy of $g_0$ with respect to the kernel $f^*$. We thus have to take into account the fact that it might be difficult to analytically evaluate the integral in \eqref{eq:new}. To this end we resort to the strategy illustrated by \citet{Fav13} and adapted from \cite{Nea00}'s algorithm 8 for DP mixtures. 
A convenient augmentation is obtained by introducing a random vector $(\bgamma_1^{(\e)},\ldots,\bgamma_r^{(\e)})$, of arbitrary size $r$, whose components are independent and identically distributed from $G_0$, and independent of $(\bgamma_{1,i}^{*},\ldots,\bgamma_{k^{(i)},i}^{*})$. The full conditional distribution \eqref{eq:fc_gamma} can then be replaced by the following.
\begin{enumerate}
\item[(c$^\prime$)] The alternative full conditional distribution for $\bgamma_i$ is given by
\begin{equation}\label{eq:neal_pred}
P(\bgamma_i \in \cdot \mid \mbox{rest}, \bgamma_1^{(\e)},\ldots,\bgamma_r^{(\e)})=\sum_{l=1}^{r}p_{l,i}^{(\e)} \mathds{1}_{\bgamma_l^{(e)}}(\cdot) + \sum_{j=1}^{k^{(i)}}p_{j,i} \mathds{1}_{\bgamma_{j,i}^{*}}(\cdot),
\end{equation}
where $p_{j,i}$ is given in (c) and, for $l=1,\ldots,r$,
$$p_{l,i}^{(e)}\propto  \frac{\alpha \kappa_{1}(u)}{r} f^*(y_i\mid \bgamma_{l,i}^{(\e)},\bx_i)^{\delta_i} S^*(y_i\mid \bgamma_{l,i}^{(\e)},\bx_i)^{1-\delta_i}.$$
\end{enumerate}

As suggested in \citet{Nea00} and \citet{Fav13}, we count the old value taken by $\bgamma_i$ as one of the components of $(\bgamma_1^{(\e)},\ldots,\bgamma_r^{(\e)})$ in case the same does not appear in $(\bgamma_{1,i}^{*} ,\ldots,\bgamma_{k^{(i)},i}^{*})$.

\subsection{Choice of L\'evy intensity}\label{sec:levy}

We next choose convenient parametric forms for the functions $G_0$ and $\rho$, appearing in the L\'evy intensity \eqref{eq:levy} which characterizes the NRMI prior for $G$.  Specifically, by focusing on the more general framework described by model M2, we assume that $G_0$ is given by the independent product of $m-1$ normal densities with mean $\mu_{0,l}$ and variance $\tau_{0,l}^2$, with $l=1,\ldots,m-1$, and an inverse Gamma density function with shape and scale parameters denoted by $q_0^{(\gamma)}$ and $q_1^{(\gamma)}$ respectively. The normal components of $G_0$ model the location parameter $\gamma_{i,1}=\mu_i$ and the regression coefficients $\gamma_{i,l}=\theta_{i,l-1}$, for $l\in\{2,\ldots,m-1\}$, the inverse Gamma density function deals with the scale factor $\gamma_{i,m}=\zeta_{i}$. 
As for the non-negative function $\rho$ we concentrate on the class defined by the inverse Gaussian process, that is we set $\rho(s)=\frac{1}{2\sqrt{\pi}}s^{-3/2}e^{-\tau s}$, where $\tau>0$, which makes $G$ distributed as the normalized inverse Gaussian (N-IG) process introduced by \cite{Lij05}. While retaining most of the tractability characterizing the more popular DP, the N-IG features an appealing clustering property which makes a richer use of the information contained in the data \citep[see][]{Lij07}. The latter property makes the N-IG a convenient choice to explore the space of data partitions for the purpose of devising an optimal stratification. The model is completed by choosing a gamma prior for $\tau$, that is  $\tau\sim\ga( q_0^{(\tau)},q_1^{(\tau)})$. Moreover, as far as model M1 is concerned, independent normal priors are adopted for the common regression coefficients $\theta_l$, that is $\theta_l\simiid\no(0,\tau_\theta^2)$, for $l=1,\ldots,m-2$. Explicit expressions for the full conditional distributions for the N-IG case, including the full conditional for the parameter $\tau$ and those for the common regression coefficients of model M1, are provided in 
\ref{sec:fc}, along with details on how to simulate from such distributions. 

\subsection{Optimal partition}\label{sec:VI}
Conditionally on a realization of the latent variables $\bgamma$, it is natural to interpret two individuals, say the $i_1$-th and $i_2$-th ones, as belonging to the same group if their latent variables take the same values, that is $\bgamma_{i_1}=\bgamma_{i_2}$. In this sense, a realization of $\bgamma$ defines a partition $\rho_{\bgamma}$ in the space $\mathcal P_n$ of all possible partitions of $n$ observations. Similarly, the posterior distribution of $\bgamma$ induces a posterior distribution on $\mathcal P_n$, the main object of our interest. In order to obtain a point estimator for $\rho_{\bgamma}$, we follow the decision theoretic approach of \citet{Wad18}, which we briefly summarize here. 
We consider a loss function $L(\rho^*, \rho)$ that measures the loss in estimating the true partition $\rho^*$ with $\rho$, where both $\rho^*, \rho \in \mathcal P_n$. When the true partition $\rho^*$ is not known, as in the case we deal with when investigating the posterior distribution of $\rho_{\bgamma}$, an estimator $\hat \rho_{\bgamma}$ can be defined as the element of $\mathcal P_n$ minimizing the expected value of the loss function $L(\rho_{\bgamma}, \rho)$ taken with respect to the posterior distribution of $\rho_{\bgamma}$, that is
\begin{equation}\label{eq:optpart}
    \hat \rho_{\bgamma} = \underset{\rho \in \mathcal P_n}{\argmin} \E [L(\rho_{\bgamma}, \rho) \mid \bY,\bdelta] = \underset{\rho \in \mathcal P_n}{\argmin} \sum_{\rho^* \in \mathcal P_n} L(\rho^*, \rho) P(\rho_{\bgamma}=\rho^* \mid \bY,\bdelta).
\end{equation}
Different choices for the loss function $L$ have been considered in the literature, such as the 0--1 loss function, the Binder loss function \citep{dahl:06,Lau07} and the variation of information loss function \citep{Wad18,Ras18}. 
Here we focus on the last one, which is given by
\begin{equation}\label{eq:VI}
    L_{\text{VI}}(\rho^*,\rho)=\sum_{i=1}^{k_n^*}\frac{n_{i\bullet}}{N}\log\left(\frac{n_{i\bullet}}{n}\right)+\sum_{i=1}^{k_n}\frac{n_{\bullet j}}{n}\log\left(\frac{n_{\bullet j}}{n}\right)-2\sum_{i=1}^{k_n^*}\sum_{j=1}^{k_n}\frac{n_{ij}}{n}\log\left(\frac{n_{ij}}{n}\right),
\end{equation}
where $k_n^*$ and $k_n$ denote the number of blocks in the partitions $\rho^*$ and $\rho$, respectively, $n_{ij}$ denotes the number of observations shared by the $i$-th block of $\rho^*$ and the $j$-th block of $\rho$, $n_{i\bullet}:=\sum_{j=1}^{k_n}n_{ij}$ and $n_{\bullet j}:=\sum_{i=1}^{k_n^*}n_{ij}$. The loss function in \eqref{eq:VI} compares the information in the two partitions, with the information shared by the two partitions. An empirical study by \citet{Wad18} suggests that, compared to the Binder loss function, $L_{\text{VI}}$ has the property of penalizing small clusters thus leading to an arguably more interpretable partition. This feature is appealing for the purpose of this work, as it is known that overstratification might lower the precision of the estimated effect of interest \citep{Des08}. 
The task of solving the optimization problem in \eqref{eq:optpart} is daunting due to the discrete nature and the cardinality of $\mathcal P_n$, very large even for moderately small values of $n$. An approximate solution is achieved by using the sample generated from the posterior distribution of $\rho_{\bgamma}$ to obtain a Monte Carlo evaluation of the expected value, and by approximating $\hat \rho_{\bgamma}$ with the partition $\rho$ which, among those generated, minimizes the approximated expectation. Alternative numerical approaches to deal with \eqref{eq:optpart} are presented in \citet{Wad18} and \citet{Ras18}. 

\section{Simulation study}\label{sec:simu}
We present an extensive simulation study to assess the ability of the proposed approach to detect a suitable stratification of the data. To this end, we consider simulation scenarios for which a natural stratification is known. This is achieved by assuming there exist three strata, which in turn we model independently so that, for all the observations belonging to the $j$-th stratum, with $j=1,2,3$, we have
\begin{equation}\label{eq:dgp}
Y_i\mid\mu_i=\mu^{(j)},\btheta_i=\btheta^{(j)},\zeta_i=\zeta^{(j)},\mathbf{x}_i  \simind  f^*(y_i\mid\mu^{(j)},\btheta^{(j)},\zeta^{(j)},\mathbf{x}_i).
\end{equation}
The data generating process is completed by specifying the parametric form of $f^*$, which in this simulation study is set equal to a type-I minimum distribution, and the effect of the covariates. The latter can be null (that is $\btheta^{(j)}=\bzero$ for every $j$), non-null and common across strata (that is $\btheta^{(1)}=\btheta^{(2)}=\btheta^{(3)}\neq \bzero$), or stratum-specific (that is $\btheta^{(j_1)}\neq \btheta^{(j_2)}$ if $j_1\neq j_2$). Such a choice leads to three distinct data generating processes that will be named model D0, D1 and D2, respectively. 
Data generated from model D0, D1 and D2, will be analyzed by considering different specifications of models M0, M1 and M2. It is apparent that there is an analogy between the set of alternative assumptions made on the effect of the covariates for the data generating processes D0, D1 and D2, and those made for the models M0, M1 and M2. Finally, $f^*$ will be set equal to \eqref{eq:densEvi}, \eqref{eq:densLog} or \eqref{eq:densN}, thus leading to a total of nine alternative models for posterior inference. Our study also considers the presence of right-censored observations and aims at investigating the robustness of the proposed methodologies when the percentage of censored observations becomes large. This will be achieved by running the same analysis on data with a varying percentage of censored observations, namely $0\%$, $10\%$, $20\%$ or $30\%$. In order to measure the similarity between true and estimated optimal partition, we consider the RAND index \citep{Ran71,Gat17}.

\begin{figure}[h!]
	\center 
	\includegraphics[width=\textwidth]{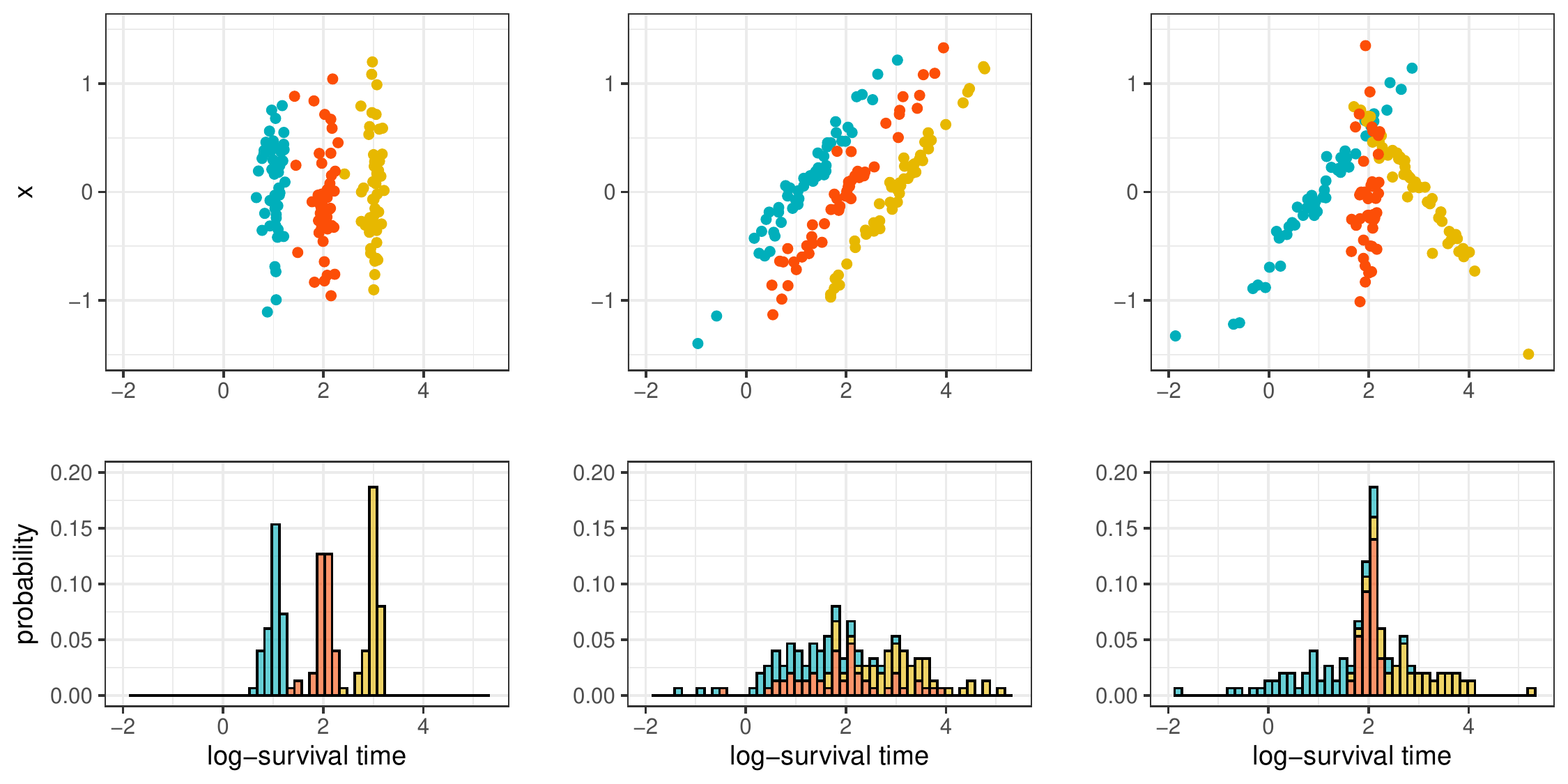}
	\caption{\label{fig:scenarios} Data simulated from a type-I minimum distribution for the log-survival times $Y_i$ and characterized by three groups, each of size 50, displayed in different colours. The bottom row displays stacked histograms for $Y_i$; the top row shows scatter plots for the joint distribution of $(Y_i,X_i)$, with the response variable on the $x$-axis and the explanatory variable on the $y$-axis. The three columns refer to three distinct data generating processes, namely model D0 (left), model D1 (middle) and model D2 (right).
	}
\end{figure}

Figure~\ref{fig:scenarios} displays an example of simulated data. For each observation, one covariate is considered, whose values are independently generated, here and for the rest of the study, from a normal distribution with null mean and variance equal to 0.25. We consider samples of $n$ log-survival times $Y_i$, composed by three equally sized groups. For the purpose of illustration, Figure~\ref{fig:scenarios} assumes $n=150$, while various values for $n$ will be considered in the rest of the study. Observations are simulated from a type-I minimum distribution \eqref{eq:densEvi}, each group being characterized by a specific set of location and scale parameters, namely $(\mu^{(1)},\zeta^{(1)})=(1,0.15)$, $(\mu^{(2)},\zeta^{(2)})=(3,0.1)$ and $(\mu^{(3)},\zeta^{(3)})=(2,0.12)$. 
Three distinct data generating processes are then obtained by setting $\theta^{(1)}=\theta^{(2)}=\theta^{(3)}=0$ (model D0), $\theta^{(1)}=\theta^{(2)}=\theta^{(3)}=-1.5$ (model D1) and $\theta^{(1)}=-1.5$, $\theta^{(2)}=1.6$, and $\theta^{(3)}=-0.1$ (model D2). 
The three data generating processes produce data displaying distinctive features and requiring different levels of model flexibility. Model D0 (left column of Figure~\ref{fig:scenarios}) generates data where both the marginal distribution of the log-survival times $Y_i$ and the joint distribution of log-survival times and covariates $(Y_i,X_i)$ display three well separated groups; as for model D1 (central column of Figure~\ref{fig:scenarios}), the three groups are well separated as far as the joint distribution of $(Y_i,X_i)$ is concerned, while there appears to be a substantial overlap between the distributions of $Y_i$ for the three groups; finally, model D2 (right column of Figure~\ref{fig:scenarios}) gives rise to data for which the three components overlap in both the marginal distribution of $Y_i$ and the joint distribution of $(Y_i,X_i)$.

\subsection{Stratification of the data}\label{sec:stratdata}
The first part of the study focuses on the ability of the proposed methods to recover the stratification composed by the three groups determined by the data generating process. To this end we considered three sample sizes $n\in\{90,150,300\}$ and, for each value of $n$, we generated 50 independent replicates from the three scenarios depicted in Figure~\ref{fig:scenarios}, all consisting of three equally sized groups of observations. We also generated data sets including censored observations: this was achieved by independently generating the censoring times from an exponential distribution with parameter tuned to obtain an overall portion of censored observations equal to 10\%, 20\% or 30\%. 
The specification of models M0, M1 and M2 is completed by setting $\alpha=1$, and the two parameters of the gamma hyperprior on 
$\tau$ equal to 1, that is $
q_0^{(\tau)}=q_1^{(\tau)}=1$. Moreover, the parameters characterizing the base measure $G_0$ are set so that $\mu_{0,1}=\bar{\bY}$, $\tau_{0,1}^2=S^2_{\bY}$ (sample variance), $q_0^{(\gamma)}=5$, $q_1^{(\gamma)}=1$, and, as far as M2 is concerned, $\mu_{0,2}=0$, $\tau_{0,2}^2=20$.
Posterior estimates for each replicate were obtained by running the Gibbs sampler for a total of $5\,000$ iterations, the first $3\,000$ of which were discarded as burn-in. Visual inspection of the traceplots of randomly selected replicates suggested a moderately good mixing and did not highlight any convergence issues.

\begin{figure}[h!]
    \centering
	\subfloat[exact\label{fig:a}]{\centering\includegraphics[width=0.5\textwidth]{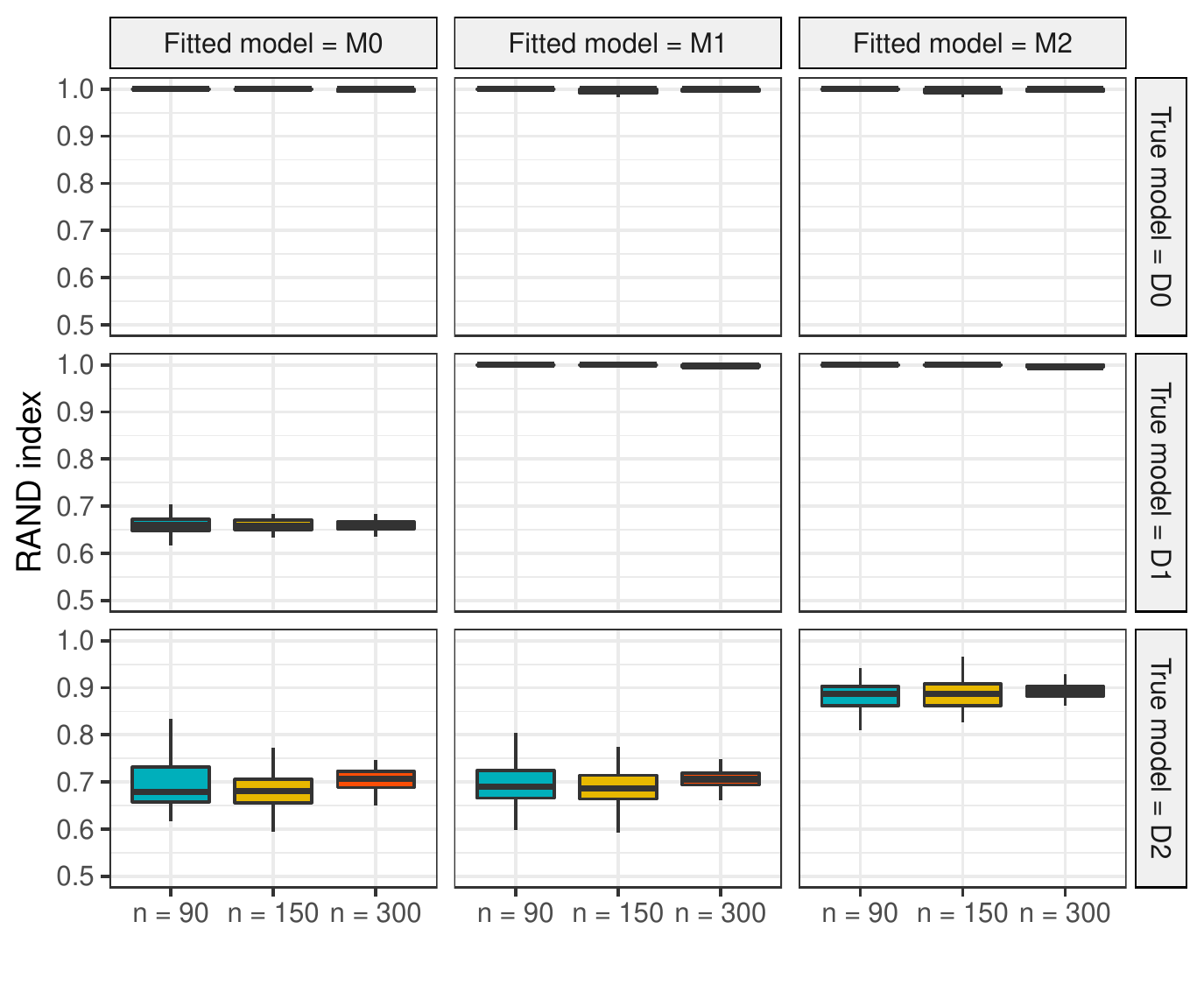}}
	\subfloat[10\% censored \label{fig:b}]{\includegraphics[width=0.5\textwidth]{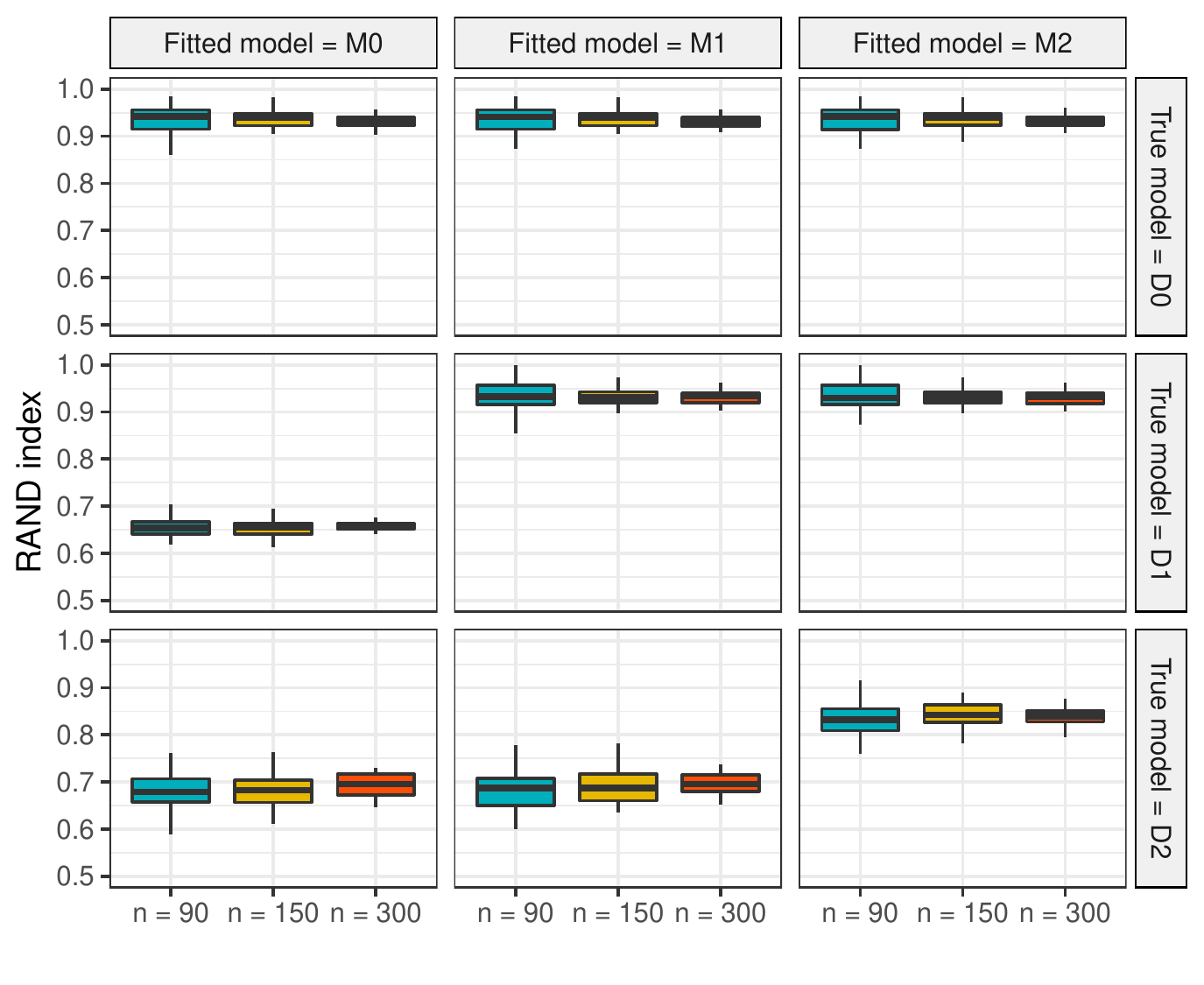}}
	\\[6pt]
	\centering
	\subfloat[20\% censored\label{fig:c}]{\includegraphics[width=0.5\textwidth]{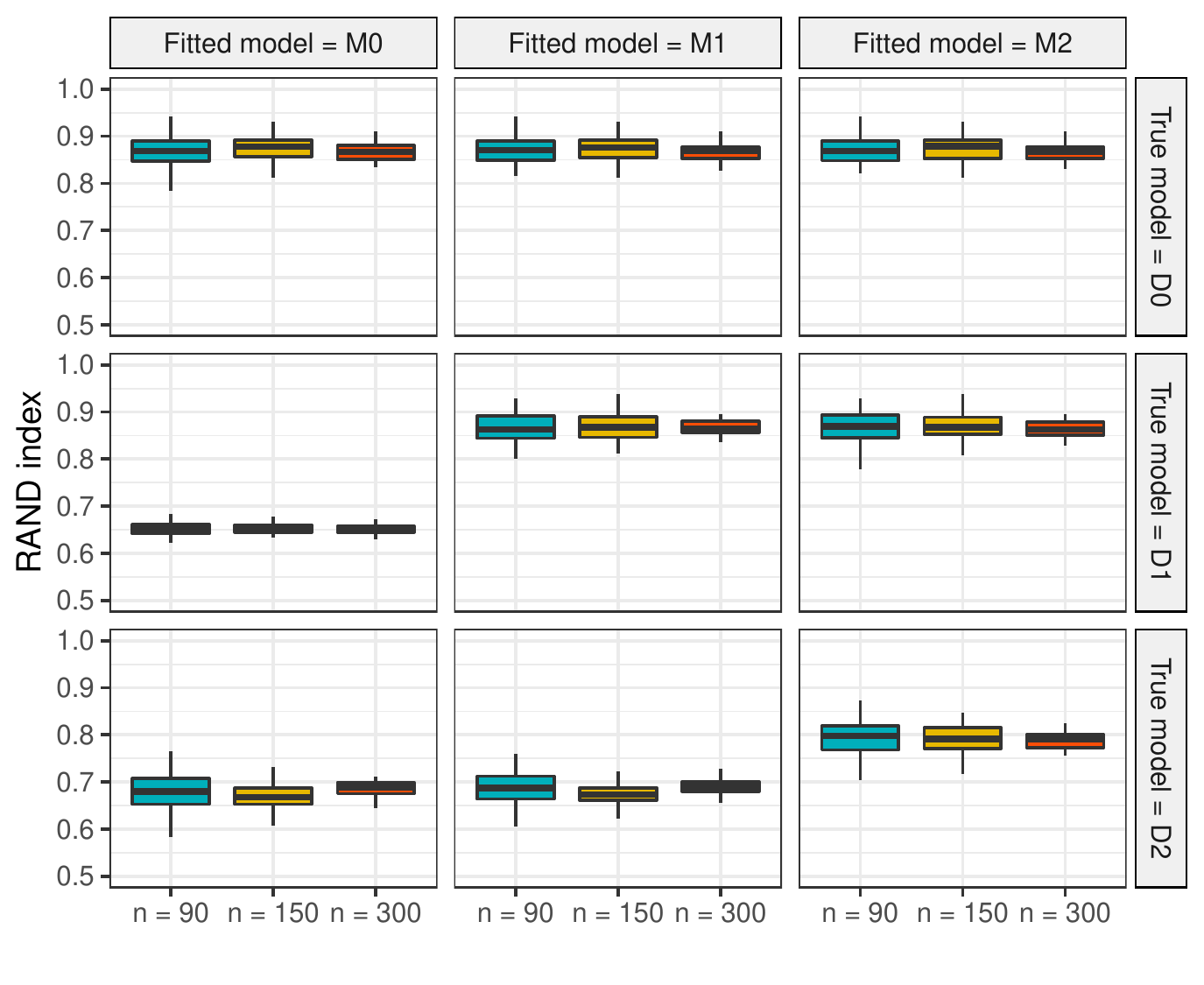}}
	\subfloat[30\% censored\label{fig:d}]{\includegraphics[width=0.5\textwidth]{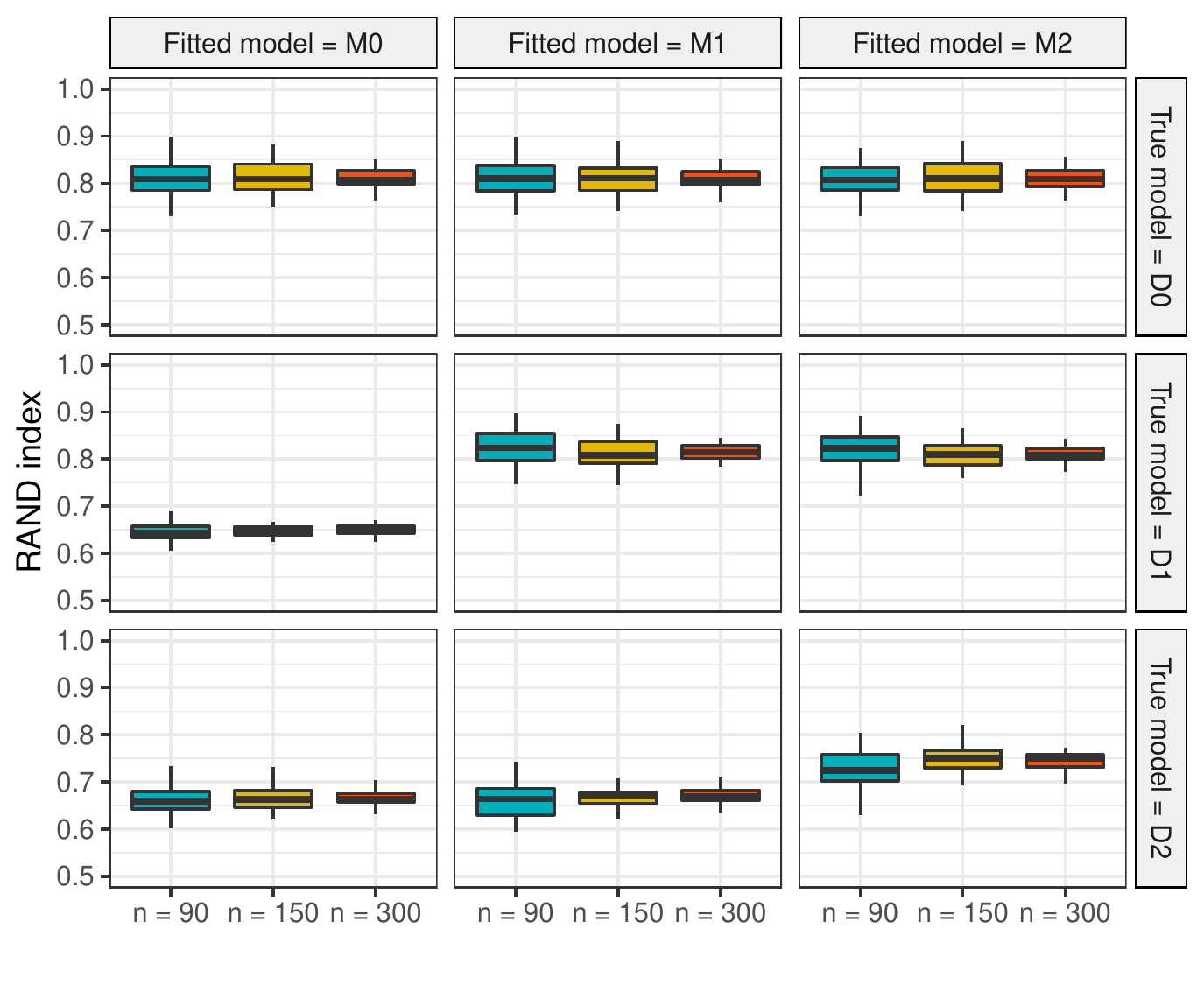}}
\caption{\label{fig:strat_1} Simulated data. RAND index, measuring the similarity between the true partition of the data and the detected optimal stratification, for different sample sizes $n$ ($n=90$ in cyan, $n=150$ in yellow, $n=300$ in red). The boxplots are obtained by analysing 50 replicates of each scenario. Both the data generating processes and the models for the log-survival times that were implemented, are defined by means of a type-I minimum kernel. The four panels refer to different percentages of right-censored observations, namely $0\%$ (panel a), $10\%$ (panel b), $20\%$ (panel c) and $30\%$ (panel d). In each panel different rows refer to different data generating processes, different columns refer to different models fitted the data.}
\end{figure}

Figure~\ref{fig:strat_1} displays boxplots of the values taken by the RAND index, measuring the similarity between the partition of the data, as determined by the data generating process, and the detected optimal stratification. More specifically, the four panels of the figure refer to data sets where all the observations are exact (Figure \ref{fig:a}), or feature a portion of right-censored observations equal to 10\%, 20\% and 30\% (Figures \ref{fig:b}, \ref{fig:c} and \ref{fig:d}, respectively).
Boxplots refer to the 50 replicates generated for each scenario and each sample size. In each of the four panels, different rows correspond to the three distinct data generating processes, namely D0, D1 and D2, while the three columns refer to the different models fitted to analyze the data, namely M0, M1 and M2, with a type-I minimum kernel for the log-survival times.
Figure \ref{fig:a} clearly shows that the three models, M0, M1 and M2, succeed in identifying the true stratification when the data are generated from model D0; model M0 fails in capturing a sensible stratification when analysing data generated from models D1 or D2, and, similarly, model M1 fails when the data generating process is M2; finally, model M2 proves able to produce a data partition close to the true one regardless of which of the three data generating processes was considered. These results are not surprising as they confirm that data generated from the simplest data generating process, that is the one with no covariates effect, are well handled by the three versions of the model we implemented; on the contrary, only the enhanced flexibility of model M2 allows us to capture a sensible stratification when the data come from the data generating process with group-specific effect of the predictors. It is more interesting to see how the same methods perform when a portion of the observations is right-censored, which is something displayed by Figures \ref{fig:b}--\ref{fig:d}. While, as far as the comparison of models M0, M1 and M2 is concerned, considerations similar to the ones made for the case of data without censoring hold, some additional comments are in order. The values taken by the RAND index are on average lower as the percentage of right-censored data becomes large. For example, when model M0 is used to analyze data generated from model D0, the value taken by the RAND index, averaged over the three sample sizes considered in the study and all the replicates, is 0.99, 0.93, 0.86 and 0.81 as the percentage of censored observations ranges in $\{0\%,10\%,20\%,30\%\}$; similarly, the average value taken by the RAND index when model M2 is used to analyze data generated from model D2, is  0.89, 0.84, 0.79 and 0.74 for the same percentages of censoring. Moreover, we observe that larger sample sizes do not appear to lead to larger values for the RAND index. On the contrary, in every considered scenario, the average value, out of 50 replicates, taken by the RAND index seems roughly constant across different sample sizes. 
What clearly changes with the sample size is the variability, around the average, of the observed values of the RAND index, with larger samples displaying smaller variability. Finally, according to our study, low values of the RAND index tend to correspond to large numbers of groups in the identified stratification. This behaviour can be appreciated by looking at Figure \ref{fig:clust_1} in \ref{sec:plots}.

While Figure \ref{fig:strat_1} refers to the case where both the data generating process and the model we implemented for the log-survival times are defined by means of a type-I minimum kernel, the same study was carried out by analysing the exact same data sets with models defined by means of a kernel which is either logistic or normal. The results, displayed in Figures \ref{fig:strat_2} and \ref{fig:strat_3} in \ref{sec:plots}, are similar and, thus, suggest that, provided that the kernel has mean zero and unit variance, its parametric form might not considerably affect the ability of the method to identify a stratification close to the correct one. 

\subsection{Stratum-specific inference}

In the second part of the study we consider the stratum-specific inference produced conditionally on the identified optimal stratification. This is produced by implementing, independently on each identified stratum, the exact same nonparametric mixture model considered to determine the stratification. At this point it is worth reporting that, in order to provide meaningful summaries of the posterior survival functions, such as posterior credible bands for the estimated survival curves, 
we resorted to the method presented by \citet{Arb16}. The idea consists in obtaining, for any time $t$, an approximation of the distribution of a random survival function $\tilde S$ evaluated at $t$, starting from the estimation of the first moments of the random variable $\tilde S(t)$. These are easily obtained from the output of a marginal algorithm, such as the one implemented in this work.

We ran two simple experiments with the goal of shedding some light on the performance of the described methods. In the first one a data set of size $n=150$  is generated from model D2 and consists of three equally-sized groups characterized by different baseline survivals as well as different covariate effects. To be more specific, we considered the same specification of model D2 that was used to generate the data analyzed in Section \ref{sec:stratdata}. The generated data set was analyzed by means of models M0, M1 and M2, with a type-I minimum kernel for the log-survival times and the same prior specifications described in Section \ref{sec:stratdata}.

\begin{figure}[h!]
	\center 
    	\includegraphics[width=\textwidth]{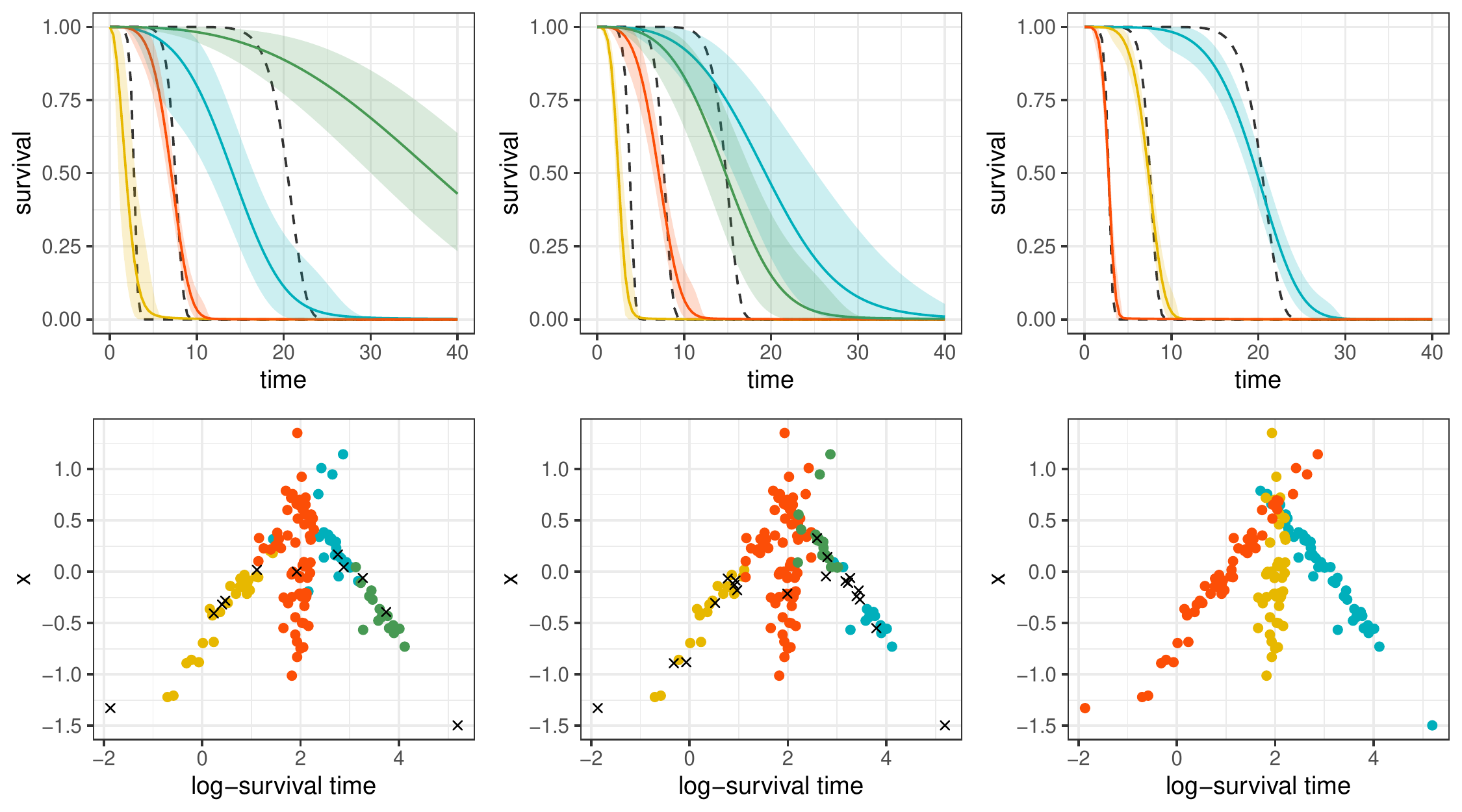}
	\caption{\label{fig:single_rep} Simulated data. Data generated from model D2 and analyzed with model M0 (left column), model M1 (middle column), and model M2 (right column). The top row displays the true stratum-specific baseline survival curves (dashed black curves),  estimated stratum-specific baseline survival curves (for all groups counting at least two individuals), along with $99\%$ posterior credible bands; the bottom row shows the scatter plots for the joint distribution of $(Y_i,X_i)$, with the response variable on the $x$-axis and the explanatory variable on the $y$-axis. Different colours correspond to different strata in the identified optimal stratification, black crosses denote individuals belonging to one-sized blocks in the estimated optimal partition and thus not assigned to any strata.
	}
\end{figure}

Figure~\ref{fig:single_rep} displays the optimal partitions detected by the three models, as well as the corresponding stratum-specific estimated baseline survival functions. The latter were obtained by considering, for each stratum, the expected value of the posterior predictive survival function. The data set is the same one depicted in the right column of Figure \ref{fig:scenarios}: a visual comparison of the colours in the scatter plot in the right column of Figure \ref{fig:scenarios} and those in the scatter plots of the bottom row of Figure \ref{fig:single_rep} makes apparent that, at least for this specific data set, model M2 is the only one able to identify a sensible stratification, composed of three groups. On the contrary, models M0 and M1 identify respectively four and five strata, that is four and five groups containing at least two observations. As a by-product, only model M2 leads to estimated stratum-specific baseline survivals which appear close to the ones used to simulate the data from model D2. While such finding is not surprising given that stratum-specific baseline survivals are estimated conditionally on the identification of different optimal stratifications, it can be appreciated that results produced by different stratification methods might lead to very different interpretations. Finally, we note that most of the data points not assigned to any of the groups (displayed with a black cross in Figure \ref{fig:single_rep}), that is data points whose block in the optimal partition is a singleton, lie close to the boundaries of the detected groups. 

The second experiment we ran aims at clarifying the performance of model M2 when the data generating process assumes constant, and possibly null, covariate effects across strata. That is we want to understand whether model M2 performs well also when its modelling flexibility is not required by the data under analysis. To this end we generated three data sets of size $n=150$ from models D0, D1 and D2, with a type-I minimum kernel for the log-survival times, respectively. The parameters of models D0, D1 and D2 are specified as in Section~\ref{sec:stratdata}. The three data sets were analyzed by means of model M2, with a type-I minimum kernel for the log-survival times and the same prior specification described in Section~\ref{sec:stratdata}. 

\begin{figure}[h!]
	\center 
	\includegraphics[width=\textwidth]{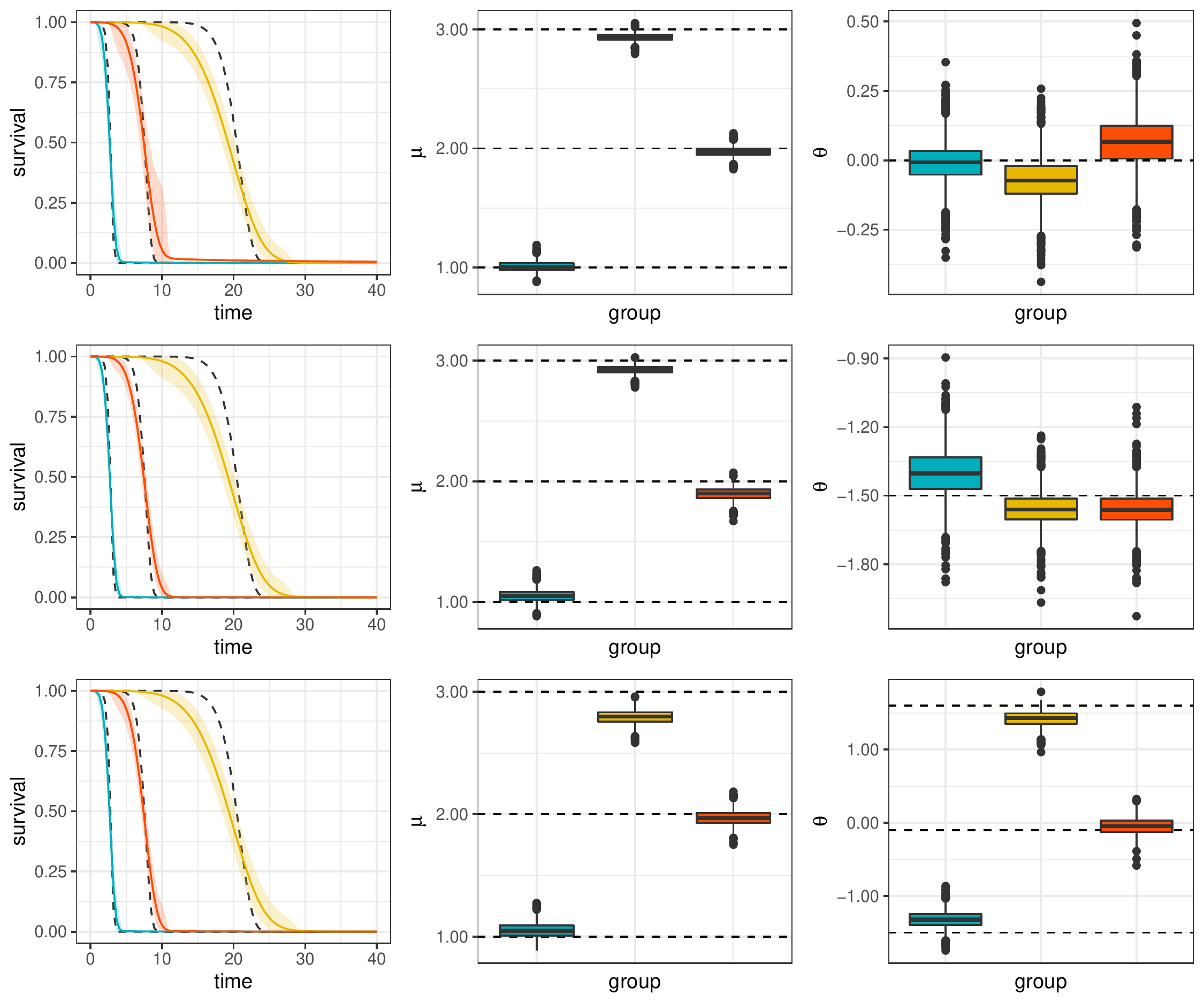}
	\caption{\label{fig:single_rep2}Simulated data. Data generated from model D0 (top row), model D1 (middle row) and model D2 (bottom row), and analyzed with model M2. Plots in the left column display true stratum-specific baseline survival curves (dashed black curves), estimated stratum-specific baseline survival curves, 
	along with $99\%$ posterior credible bands. The other two columns display boxplots of the posterior distribution for the location parameter (middle column) and regression coefficient (right column) of the identified groups, compared to the true stratum-specific values (dashed black horizontal lines) for the same parameters.
	}
\end{figure}

Our analysis resulted in the correct identification of a stratification composed by three strata for the three considered data sets. Figure~\ref{fig:single_rep2} displays stratum-specific posterior inference in the three cases. Specifically, it shows the three estimated baseline survivals, as well as the boxplots for the stratum-specific posterior distributions of the parameters $\mu$ and $\theta$. It is interesting to observe that, when a group-specific covariate effect is not required by the data generating process (rows 1 and 2 in Figure \ref{fig:single_rep2}), model M2 leads to very similar posterior distributions for the regression coefficients characterizing the three groups. Overall, whether the true covariate effect is group-specific or not does not seem to affect the ability of the model to identify a sensible stratification of the data and to capture the shape of the baseline survivals of each stratum.

\section{Real data analysis}\label{sec:illustration}

\begin{figure}[h!]
	\centering 
	\subfloat{\includegraphics[width=0.5\textwidth]{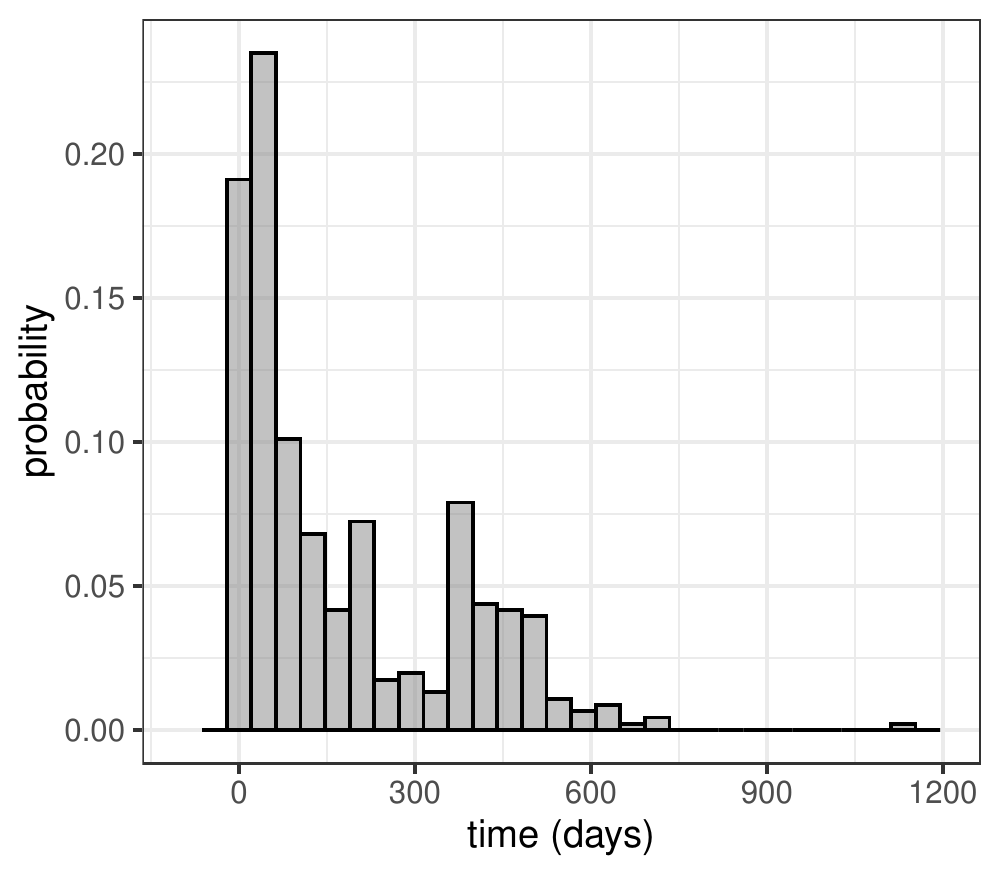}}
	\subfloat{\includegraphics[width=0.5\textwidth]{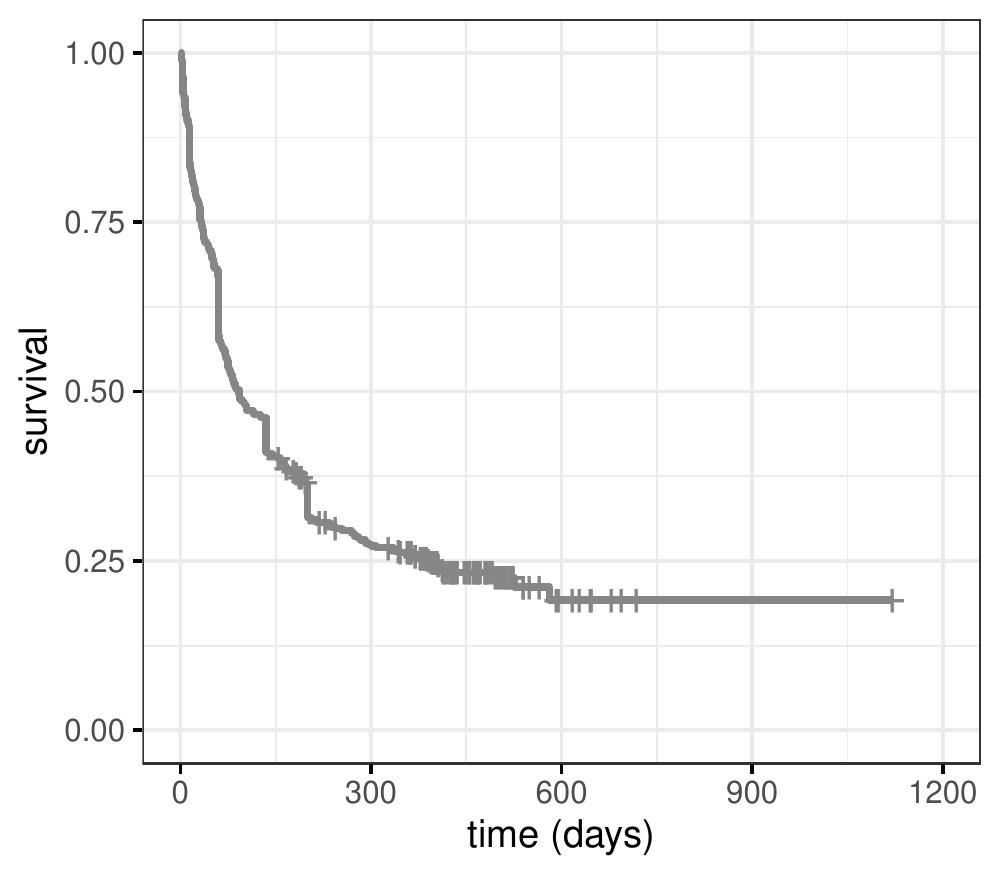}}
	\caption{\label{fig:data}UIS data set. Left panel: histogram of the times-to-relapse; right panel: Ka\-plan-Me\-ier estimate of the survival curve. 
	}
\end{figure}

We analyze data from the University of Massachusetts AIDS Research Unit IMPACT Study (UIS) \citep[see][]{McC95,Hos98}, a collaborative research project comprised of two concurrent randomized trials of residential treatment for drug abuse. The time-to-event variable is the time to return to drug use, henceforth referred to as time-to-relapse, measured in days from admission. Here we consider a data set of size $n=455$, consisting of all individuals in the study who display a positive time-to-relapse after the admission. For each individual time-to-event, two covariates are available, namely age of the subject and treatment duration, which, for the purpose of the analysis, are shifted so that their empirical distributions are centered around zero. The data set features $110$ right-censored observations (24\% of the total), corresponding to the case of individuals for which the relapse has not happened before the end of the study or the moment they withdrew from the study. 
Figure~\ref{fig:data} displays a histogram of the data set, including both exact and right censored observations, and the Kaplan-Meier estimate of the overall survival function. 

We are interested in identifying homogeneous subgroups of individuals sharing the same covariates effect and who might be consistently modelled as part of the same stratum in a stratified accelerated life model. To this end and motivated by the simulation study presented in Section \ref{sec:simu}, we assume model M2, as described in \eqref{eq:bnpmm2}. We then consider the three kernels presented in Section \ref{sec:model} and compare their predictive ability by means of the log-pseudo marginal (LPML) method of \cite{Gei79}. The results are displayed in Table \ref{tab:bestmodel} and indicate that the model built starting from the type-I minimum kernel has the best predictive ability, corresponding to the largest value of the LPML. This finding is corroborated by the values taken by the Watanabe–Akaike information criterion (WAIC) \citep{Wat10}, for which the model with type-I minimum kernel takes the smallest value.
\begin{table}[h!]
\centering
\begin{tabular}{lll}
kernel        & LPML     & WAIC   \\ \hline
type-I minimum & -268.261 & 887.031  \\
logistic & -274.043 & 906.081 \\
normal & -347.230 & 1072.023\\
\end{tabular}
\caption{UIS data set. Comparison of predictive ability, via LPML and WAIC, of three versions of model M2 with N-IG mixing measure, obtained by choosing type-I minimum, logistic and normal kernel, for the logarithm of the variable time-to-relapse.\label{tab:bestmodel}}
\end{table}

On the basis of these results, we next display the posterior inference we carried out based on the model with type-I minimum kernel for the log-times-to-relapse. The parameters of the base measure $G_0$ are specified by setting $\mu_{0,1}=\bar{\bY}$, $\tau_{0,1}^2=S^2_{\bY}$, $\mu_{0,2}=\mu_{0,3}=0$, $\tau_{0,2}^2=\tau_{0,3}^2=20$, $q_0^{(\gamma)}=5$ and $q_1^{(\gamma)}=1$. Finally, as in Section \ref{sec:simu}, $\alpha$ is set equal to 1 and the gamma hyperprior on  $\tau$ is specified by setting $q_0^{(\tau)}=q_1^{(\tau)}=1$. We ran the model for $30\,000$ iterations,  $10\,000$ of which were discarded as burn-in, and we thinned the chain keeping one value every ten realizations. Visual investigation of the traceplots indicates moderately good mixing of the chain and does not provide any indication against convergence. More information on the quality of the posterior sample is provided in \ref{sec:app_diag}. The optimal partition was chosen, among those visited by the Markov chain Monte Carlo algorithm, by adopting the variation of information criterion described in Section \ref{sec:VI}. 
Five strata were identified in the data set: the numerosity of each stratum and the corresponding composition of exact and right-censored observations are reported in Table \ref{tab:frequencies}, along with the labels and the colours we assigned to the strata for the rest of the section. A comparison with the stratification obtained by an alternative version of the same model, defined by replacing the N-IG mixing measure with a DP with coinciding prior expected number of strata, is presented in \ref{sec:app_DP}.
\begin{table}[h!]
\centering
\begin{tabular}{lcccccc}
stratum & 1 (cyan) & 2 (yellow) & 3 (red) & 4 (green) & 5 (gray) &  total \\
\hline
exact & 27 & 85 & 70 & 147 & 16 & 345 \\ 
censored & 0 & 0 & 0 & 8 & 102 & 110\\
\hline
total  & 27  & 85 & 70 & 155 & 118 & 455\\
\end{tabular}
\caption{UIS data set. Numerosity of the five strata in the estimated optimal stratification of the data, along with number of exact and right-censored observations. The first row assigns a label and a colour to each stratum. \label{tab:frequencies}}
\end{table} 
\begin{figure}[h!]
	\centering 
    \subfloat{\includegraphics[width=1\textwidth]{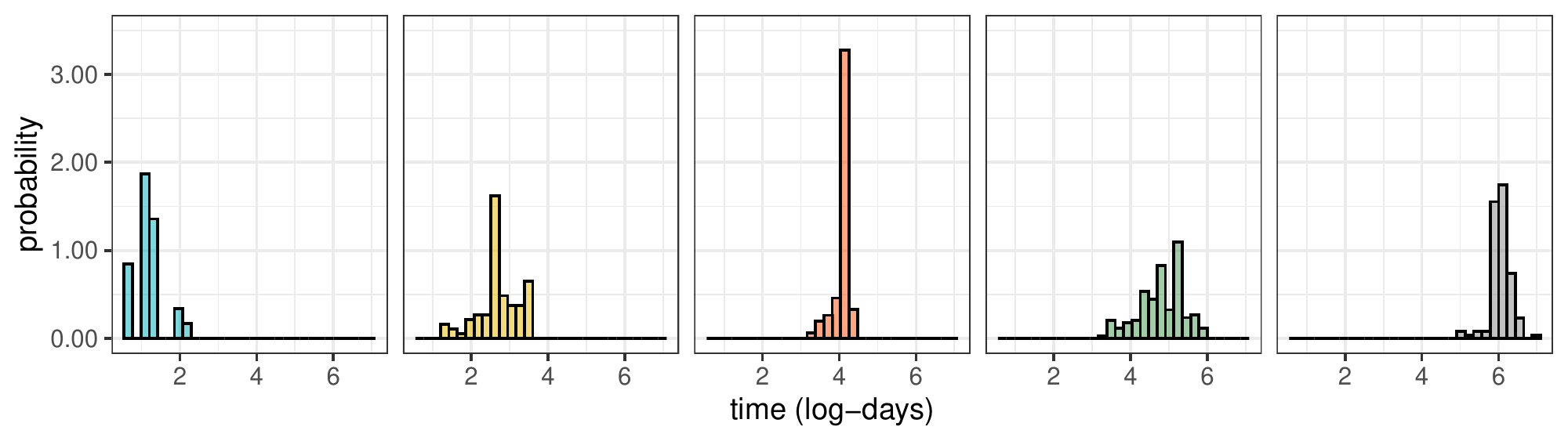}}\\
	\subfloat{\includegraphics[width=0.333\textwidth]{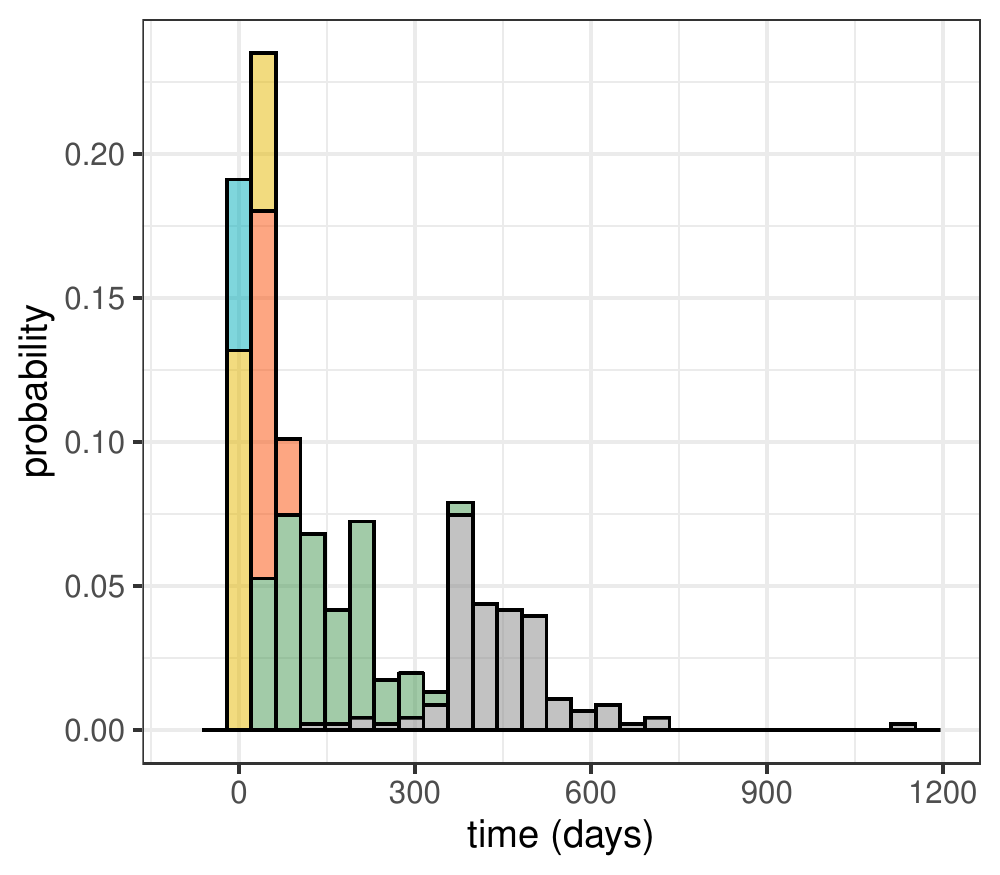}}
	\subfloat{\includegraphics[width=0.333\textwidth]{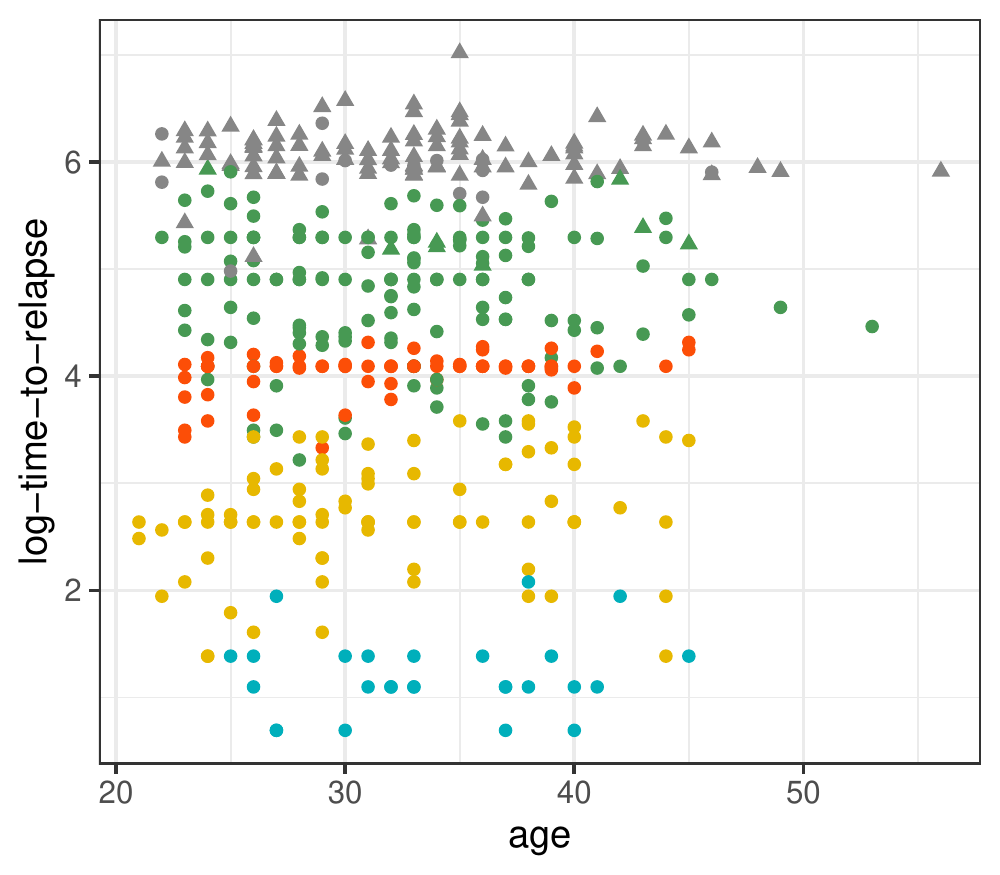}}
	\subfloat{\includegraphics[width=0.333\textwidth]{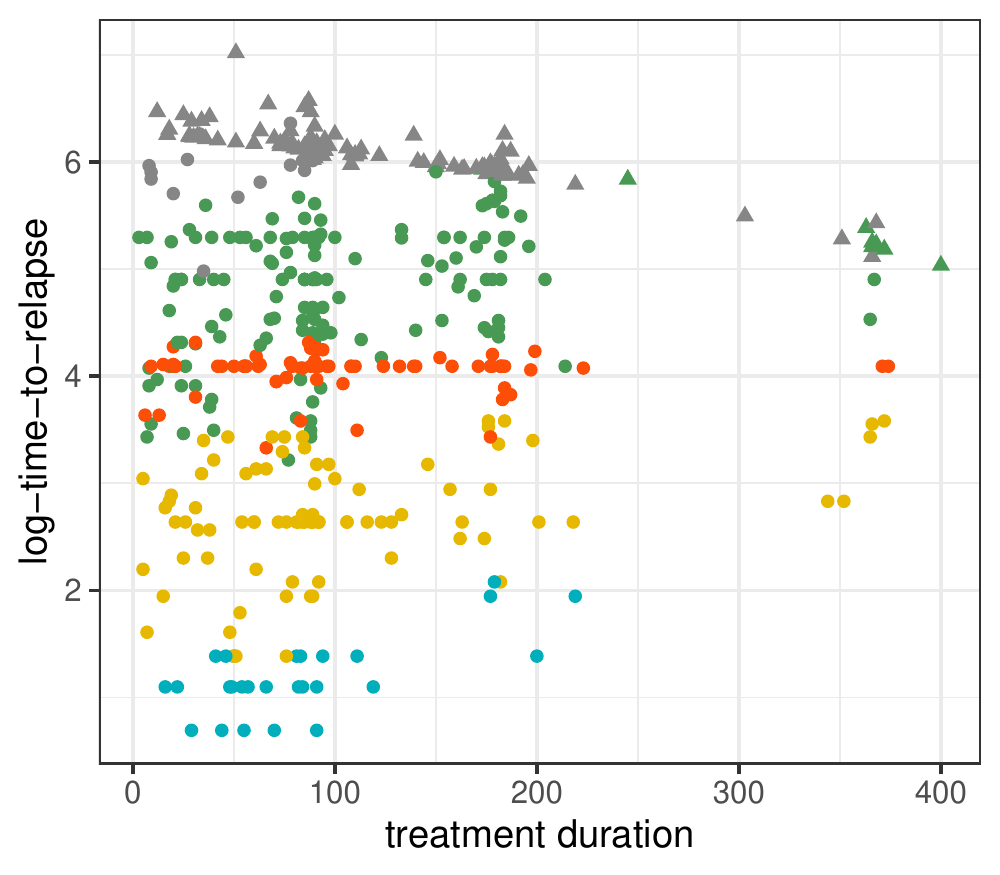}}
	
	\caption{\label{fig:data2}UIS data set. Top row: stratum-specific histograms of the variable time-to-relapse. Bottom row, left panel: stacked histogram of the variable time-to-event; bottom row, middle panel: scatter plot of age and the logarithm of time-to-relapse; bottom row, right panel: scatter plot of treatment duration and the logarithm of time-to-relapse. Different colours indicate membership to different blocks of the estimated optimal partition; in the scatter plots, circles and triangles denote exact and right-censored observations respectively. 
	}
\end{figure} The probability histograms for the times-to-relapse of each stratum are displayed in the top row of Figure~\ref{fig:data2}. The left panel of the bottom row of the same figure presents a stacked histogram of observed times-to-relapse, accounting for all strata and with bins coloured according to membership of the individuals to the five blocks composing the estimated optimal partition. The same rationale is used to colour the dots in the scatter plots of the log-times-to-relapse against the predictors age and treatment duration (middle and right panels of the bottom row of Figure \ref{fig:data2}). Two strata contain right-censored observations, 93\% of which belong to stratum 5, coloured in gray in Figure \ref{fig:data2}. This indicates that the majority of individuals in group 5 did not relapse before the end of the study, thus showing later or, possibly, no relapse.

Conditionally on the identified stratification, the same specification of model M2 was re-run independently for each stratum, to produce stratum-specific posterior inference. 
\begin{figure}[h!]
	\centering 
		\subfloat{\includegraphics[width=0.5\textwidth]{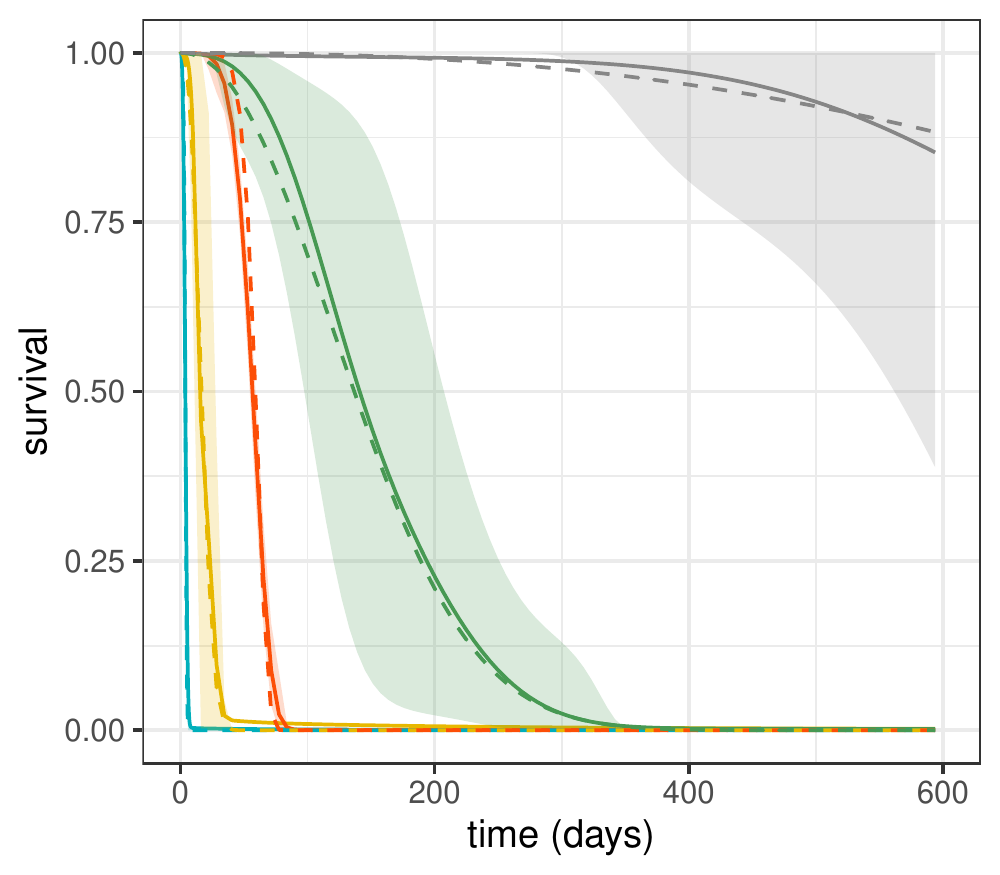}}
	\subfloat{\includegraphics[width=0.5\textwidth]{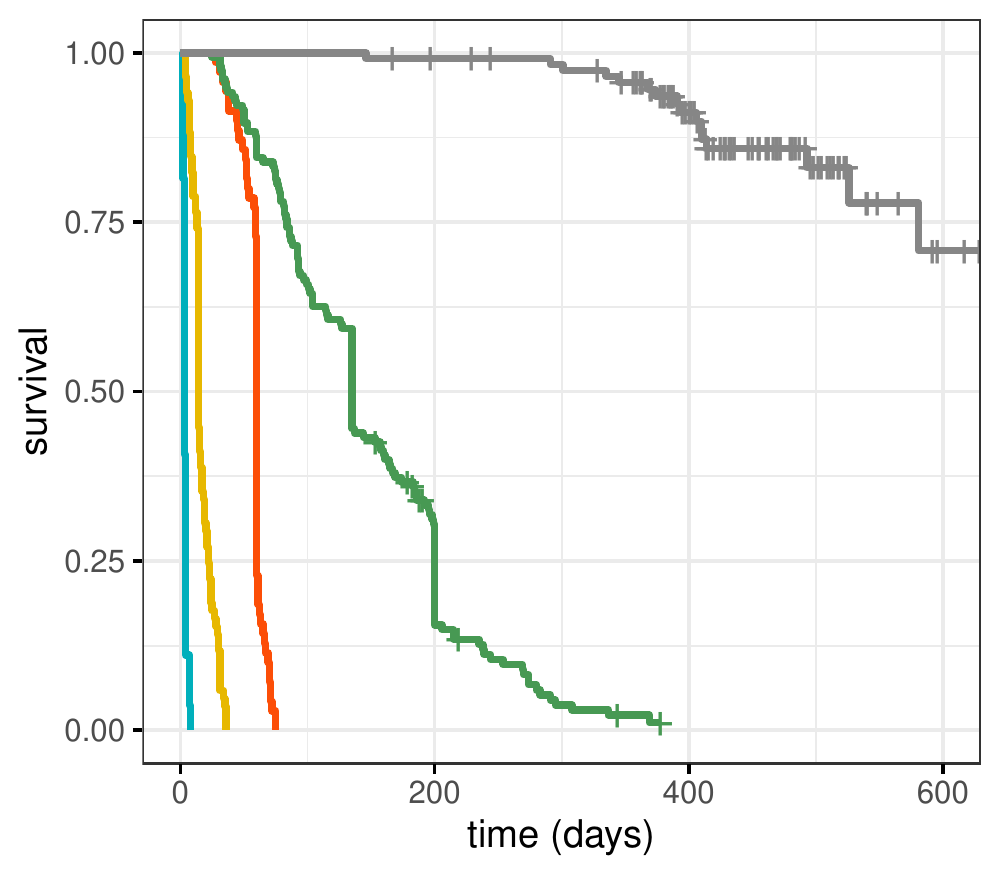}}	
	\caption{\label{fig:surv} UIS data set. 
	Left panel: stratum-specific estimated baseline survival curves with $95\%$ posterior credible bands, dashed lines correspond to the cluster specific maximum likelihood estimates; right panel: stratum-specific Kaplan-Meier estimates of the survival curves. 
	}
\end{figure}
The left panel of Figure \ref{fig:surv} displays the stratum-specific estimated baseline survival curves, that is the expected values of the corresponding posterior predictive survivals, along with $95\%$ quantile-based posterior credible bands, obtained by resorting to the method described in \citet{Arb16}. The five identified groups appear to be characterized by well-separated baseline survivals. The group coloured in gray, for which a large portion of the observations is right-censored, displays a slower decrease of the survival curve and larger credible bands. Posterior estimated survivals are compared with stratum-specific maximum likelihood estimates for the type-I minimum model for the log-times-to-relapse, and with the corresponding Kaplan-Meier estimates, displayed in the right panel of the same figure.

\begin{figure}[h!]
	\center 
	\includegraphics[width=\textwidth]{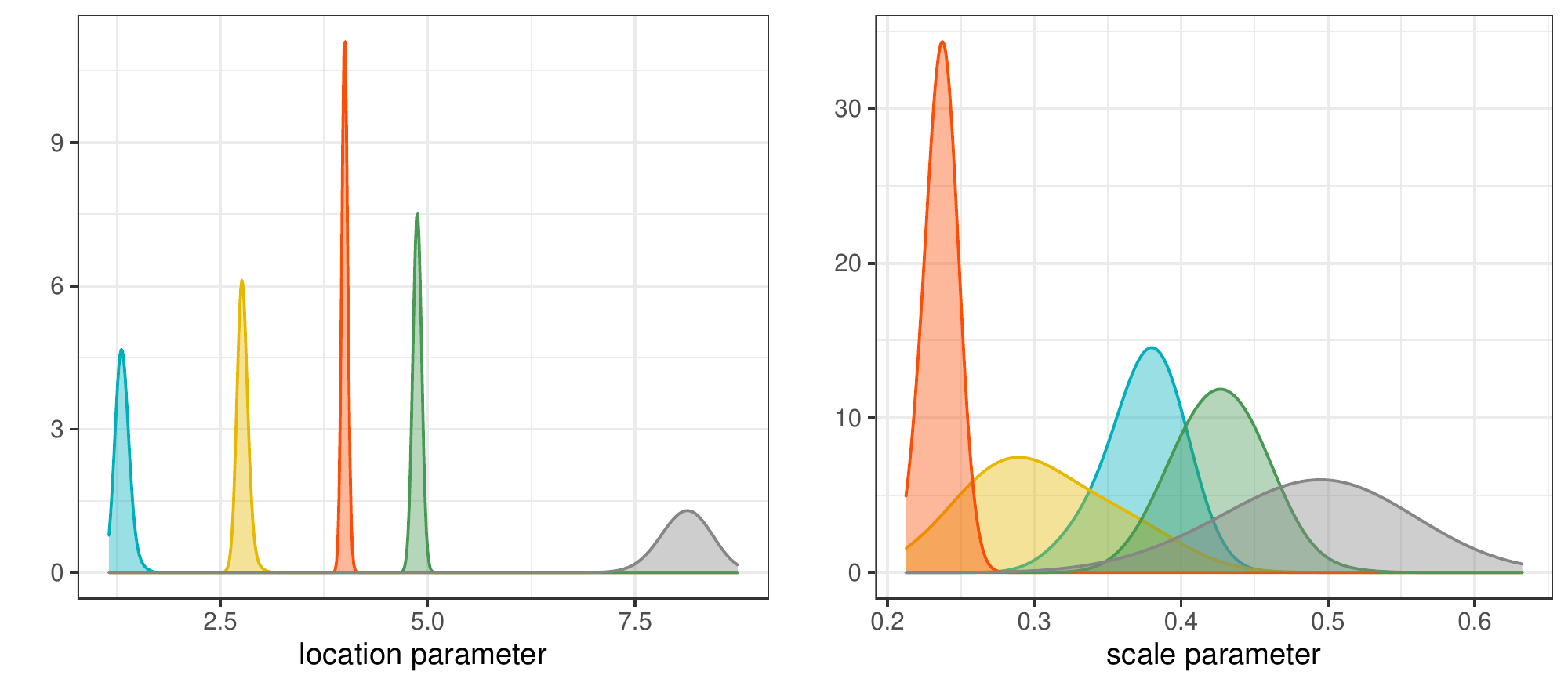}
	\caption{\label{fig:locscale} UIS data set. Stratum-specific posterior distributions for the location parameter $\mu$ (left panel) and the scale parameter $\zeta$ (right panel). Different colours correspond to different strata in the estimated partition of the data.}
\end{figure}

Our analysis also produced stratum-specific posterior estimates for location and scale parameters $\mu$ and $\zeta$, and for the regression coefficients of age and treatment duration. The posterior distributions of $\mu$ and $\zeta$ are displayed in Figure~\ref{fig:locscale}. As already highlighted by the stratum-specific estimated survival functions, Figure~\ref{fig:locscale} shows that the baseline survivals of the five groups have well-separated locations, while the posterior distributions of the stratum-specific scale parameters display a substantial overlap.   
The group in gray stands out as the posterior mean 
of both location and scale parameters is larger than those characterizing the other groups.

\begin{table}[h!]
\centering
\begin{tabular}{c cc cc}
  \multirow{2}{*}{stratum}                              
        & \multicolumn{2}{c}{age}
        & \multicolumn{2}{c}{treatment duration}\\
        \cline{2-5}  & median & 95\% c.i. & median &95\% c.i.  \\ \hline
        1 &0.072 & (-0.154 ; 0.208) &-0.382& \textbf{(-0.572; -0.250)}\\
        2 &-0.080 & (-0.209 ; 0.030) &-0.113& \textbf{(-0.261; -0.001)}\\
        3 &-0.052 & (-0.112 ; 0.000) &0.007&  (-0.045;  0.057)\\
        4 &0.030 & (-0.051 ; 0.125) &-0.317 &\textbf{(-0.436; -0.176)}\\
        5 &-0.753 & \textbf{(-1.037 ;-0.426)} &-0.843 &\textbf{(-1.063; -0.617)}\\
\end{tabular}
\caption{UIS data set. Posterior median, and corresponding 95\% quantile-based posterior credible intervals (95\% c.i.) for the regression parameters for the five identified strata. Intervals in bold font do not contain zero.\label{tab:coeff}}
\end{table}

Finally, Table~\ref{tab:coeff} reports the estimated posterior medians, along with 95\% quantile-based posterior credible intervals, for the regression coefficients of the predictors age and treatment duration, for the five identified strata. The effect of the age on the response variable appears significant only for stratum 5, for which the effect is negative, thus indicating that, for individuals belonging to that group, older age contributes in slowing the expected time-to-relapse. Treatment duration appears significant for strata 1, 2, 4 and 5, groups for which the effect is negative, which confirms that a longer duration of the treatment results in a retarded expected time-to-relapse.

\section{Conclusions}\label{sec:discussion}

One of the goals of this work is to describe, in as much generality as possible, a Bayesian nonparametric framework for stratifying survival data. The method we propose differs from standard procedures for stratification as strata are not determined on the basis of the values taken by one covariate, or a combination of covariates, but on the basis of the estimated effect of the covariates and the observed survival times. To this end we resorted to mixture models with mixing measure belonging to the rich class of NRMIs. The discreteness of NRMIs is the starting point for us to devise a procedure to find an optimal stratification, where optimality holds in the sense that the stratification we select minimizes the variation of information loss function on the space of all possible stratifications. Such procedure does not take into account the goodness-of-fit of the resulting stratified model to the data, for which other criteria should be used, such as, for example, the LPML method considered in Section \ref{sec:illustration}. Although most of the examples presented here are based on the type-I minimum kernel, we showed that other kernels belonging to the log-location-scale family of distributions can be easily implemented. In our experience, performance is somehow sensible to the prior specification of the model, as in many model-based clustering algorithms \citep[e.g.][]{nieto&contreras:13}. Thus a conservative use of the inferred stratification is advised. All models were implemented in C++, interfaced with the R environment, and the code of the main routines is available at the GitHub repository of the first author.

\section{Acknowledgments}
The first and the third authors are grateful to the \emph{DEMS Data Science Lab} for supporting this work by providing computational resources. The second author acknowledges support from \emph{Asociaci\'on Mexicana de Cultura A.C.}--Mexico.

\bibliographystyle{abbrvnat}
\bibliography{refs.bib}

\clearpage
\appendix

\section{Full conditional distribution for the N-IG model}\label{sec:fc}

When considering a N-IG mixing random probability measure, the Laplace exponent $\psi$ and the functions $\kappa_{n_j}$ become 
$$\psi(u)=\sqrt{u+\tau}-\sqrt{\tau}\quad \mbox{ and } \quad \kappa_{n_j}(u)=\frac{\Gamma\left(n_j-1/2\right)}{2\sqrt{\pi}}(u+\tau)^{1/2-n_j}.$$
The joint density of $(\bgamma, \bdelta, \bY, U)$, conditionally on $\bX$, provided in Proposition~\ref{prop:posterior}, then becomes
\begin{multline}
\nonumber
\left(\frac{\alpha}{2\sqrt{\pi}}\right)^k\frac{u^{n-1}}{\Gamma(n)} \e^{-\alpha(\sqrt{u+\tau}-\sqrt{\tau})} (u+\tau)^{k/2-n}\\
\times\prod_{j=1}^{k} \Gamma\left(n_j-\frac{1}{2}\right) \left(\prod_{l=1}^{m-1}\no(\gamma_{j,l}^*\mid \mu_0,\tau_0)\right)\iga(\gamma_{j,m}^*\mid q_0^{(\gamma)},q_1^{(\gamma)}) \d \bgamma_j^*\notag\\
 \times\prod_{i\in C_j} f^*(y_i\mid \bgamma_j^*,\bx_i)^{\delta_i}S^*(y_i\mid \bgamma_j^*,\bx_i)^{1-\delta_i}\notag,
\end{multline}
where $\bgamma_j^*=(\gamma_{j,1}^*,\ldots,\gamma_{j,m}^*)$. The resulting full conditional distributions for the random elements $(\bgamma,\alpha,U)$ and the hyperparameter $\tau$ involved in the N-IG specification, are as follows.

\begin{enumerate}

\item[(a)] The full conditional of $\alpha$ is given by
\begin{equation}\label{eq:fc_c_NIG}
f(\alpha\mid\rest)=\ga(\alpha\mid q_0^{(\alpha)}+k,q_1^{(\alpha)}+\sqrt{u+\tau}-\sqrt{\tau}).
\end{equation}

\item[(b)] The full conditional of $U$ is given, for any $u$ in $\mathds{R}^+$, by 
\begin{equation}\label{eq:fc_u_NIG}
f(u\mid \rest)\propto u^{n-1}\e^{-\alpha\,\sqrt{u+\tau}}(u+\tau)^{k/2-n}.
\end{equation}

\item[(c$^\prime$)] The full conditional distribution for each parameter vector $\bgamma_i$, $i=1,\ldots,n$, is given by \eqref{eq:neal_pred} where, for $j=1,\ldots,k^{(i)}$,
\begin{equation}\label{eq:pred_old_NIG}
p_{j,i}\propto \left(n_j-\frac{1}{2}\right)f^*(y_i\mid \bgamma_{j,i}^*,\bx_i)^{\delta_i}S^*(y_i\mid \bgamma_{j,i}^*,\bx_i)^{1-\delta_i},
\end{equation}
while, for $l=1,\ldots,r$,
\begin{equation}\label{eq:pred_new_NIG}
p_{l,i}^{(e)}\propto  \frac{\alpha\sqrt{u+\tau}}{2 r} f^*(y_i\mid \bgamma_{l,i}^{(\e)},\bx_i)^{\delta_i} S^*(y_i\mid \bgamma_{l,i}^{(\e)},\bx_i)^{1-\delta_i}.
\end{equation}

\item[(d)] The full conditional of $\tau$ has support $\mathds{R}^+$ and is given by
\begin{equation}\label{eq:fc_tau_NIG}
f(\tau\mid\rest)\propto\pi(\tau)\e^{-\alpha(\sqrt{u+\tau}-\sqrt{\tau})}(u+\tau)^{k/2-n},
\end{equation}
where $\pi(\tau)$ is the prior distribution of $\tau$.

\end{enumerate}

The implementation of model M1, for which $\bgamma_i=(\mu_i,\zeta_i)$ for any $i=1,\ldots,n$, requires to update the vector of common regression coefficients $\btheta$. 
This can be done one component per time, that is we can update each $\theta_l$ at a time, for $l=1,\ldots,m-2$ from the following distribution. 
\begin{enumerate}
\item[(f)] The full conditional of $\theta_l$ is given by
\begin{equation}\label{accelerate}
f(\theta_l\mid\rest)\propto \pi(\theta_l)\prod_{i=1}^n f^*(y_i\mid	\bgamma_i,\btheta,\bx_i)^{\delta_i}S^*(y_i\mid	\bgamma_i,\btheta,\bx_i)^{1-\delta_i}.
\end{equation}
\end{enumerate}

As for the update of $U$ and $\tau$, we resort to a random walk Metropolis--Hastings algorithm. Specifically, for the update of $U$ and $\tau$, as suggested by \citet{Fav13}, we take the log-transformation of the quantities and update them by using a normal proposal density. The variances of the proposal distributions are tuned to attain optimal acceptance rates \citep[see][]{Rob97}. Finally, in order to improve the mixing of the chain, we implemented a reshuffling step, as suggested, e.g., by \citet{Ish01}. This was done by means of a random-walk Metropolis--Hastings with proposal joint density which is the product of independent normals (for the $m$-th component $\gamma_{j,m}^*$ a log-transformation was considered).

\section{Proof of Proposition~\ref{prop:posterior}}\label{sec:proof1}
We display here the proof of Proposition~\ref{prop:posterior}, straightforward adaptation of the proof of \citet{Jam09} to the case of possibly right-censored observations.

\begin{proof} 
We define $\X^*=\X\setminus \{\d \bgamma_1^*,\ldots,\d \bgamma_k^*\}$,
and write the joint distribution of $(\bgamma,\bY,\bdelta)$, conditionally on $\bX$, as 
\begin{align*}\
\E_\mt & \left[\prod_{i=1}^n f^*(y_i\mid \bgamma_i,\bx_i)^{\delta_i}S^*(y_i\mid \bgamma_i,\bx_i)^{1-\delta_i} \frac{\mt(\d \bgamma_i)}{\mt(\X)}\right]\\
&=\E_\mt\left[\mt(\X)^{-n} \prod_{j=1}^k \mt(\d \bgamma_j^*)^{n_j} \prod_{i\in C_j}  f^*(y_i\mid \bgamma_j^*,\bx_i)^{\delta_i}S^*(y_i\mid \bgamma_j^*,\bx_i)^{1-\delta_i} \right]\\
&=\frac{\mathcal{A}_k}{\Gamma(n)}\int_0^\infty t^{n-1}\E_\mt\left[\e^{-t\mt(\X)}\prod_{j=1}^{k} \mt(\d \bgamma_i^*)^{n_j}\right]\d t\\
&=\frac{\mathcal{A}_k}{\Gamma(n)}\int_0^\infty t^{n-1}\E_\mt\left[\e^{-t\mt(\X^*)}\prod_{j=1}^{k} \e^{-t\mt(\d \bgamma_j^*)}\mt(\d\bgamma_i^*)^{n_j}\right]\d t,\end{align*}
where we have used the notation 
\begin{equation}\label{eq:short}
\mathcal{A}_k:=\prod_{j=1}^k \prod_{i\in C_j} f^*(y_i\mid \bgamma_j^*,\bx_i)^{\delta_i}S^*(y_i\mid \bgamma_j^*,\bx_i)^{1-\delta_i}.
\end{equation}
Exploiting the independence of increments of $\mt$, we rewrite the last expression as
\begin{align*}
\frac{\mathcal{A}_k}{\Gamma(n)} &\int_0^\infty t^{n-1}\E_\mt\left[\e^{-t\mt(\X^*)}\right]\prod_{j=1}^{k} \E_\mt\left[\e^{-t\mt(\d \bgamma_j^*)}\mt(\d\bgamma_i^*)^{n_j}\right]\d t\\
&=\frac{\mathcal{A}_k}{\Gamma(n)} \int_0^\infty t^{n-1}\E_\mt\left[\e^{-t\mt(\X^*)}\right]\prod_{j=1}^{k} (-1)^{n_j}\frac{\partial^{n_j}}{\partial t^{n_j}} \E_\mt\left[\e^{-t\mt(\d \bgamma_j^*)}\right]\d t\\
&=\frac{\mathcal{A}_k}{\Gamma(n)}\int_0^\infty t^{n-1}\e^{-\alpha G_0(\X^*)\psi(t)}\prod_{j=1}^{k}\left(-1\right)^{n_j+1} \e^{-\alpha G_0(\d \bgamma_j^*)\psi(t)} \alpha G_0(\d \bgamma_j^*)\frac{\partial^{n_j}}{\partial t^{n_j}} \psi(t) \d t,
\end{align*}
where the last expression is obtained by applying Fa\`a di Bruno formula for multiple derivatives of composite functions and considering only terms of first order due to the diffuseness of $G_0$. Thus we can write the joint distribution of $(\bgamma,\bY,\bdelta)$ conditionally on $\bX$, as
\begin{multline*}
\frac{(-1)^{n+k}\alpha^k \,\mathcal{A}_k}{\Gamma(n)} \int_0^\infty t^{n-1}\e^{-\alpha G_0(\X)\psi(t)}\prod_{j=1}^{k}G_0(\d \bgamma_j^*) \frac{\partial^{n_j}}{\partial t^{n_j}} \psi(t) \d t\\
=\frac{(-1)^{n+k}\alpha^k\,\mathcal{A}_k}{\Gamma(n)} \int_0^\infty t^{n-1}\e^{-\alpha\psi(t)}\prod_{j=1}^{k} \frac{\partial^{n_j}}{\partial t^{n_j}} \psi(t) \d t\,\prod_{j=1}^k G_0(\d \bgamma_j^*).
\end{multline*}
We further observe that 
\begin{equation*}
\frac{\partial^{n_j}}{\partial t^{n_j}}\psi(t)=(-1)^{n_j+1}\int_{0}^{\infty}\e^{-s t}s^{n_j}\rho(s) \d s=(-1)^{n_j+1} \kappa_{n_j}(t).
\end{equation*}
This implies that the conditional joint distribution of the joint distribution of $(\bgamma,\bY)$ becomes 
\begin{align*}
\frac{\alpha^k \,\mathcal{A}_k}{\Gamma(n)}\int_0^\infty t^{n-1}\e^{-\alpha\psi(t)}\prod_{j=1}^{k} \kappa_{n_j}(t) \d t\,\prod_{j=1}^k G_0(\d \bgamma_j^*).
\end{align*}
The introduction of a convenient auxiliary random variable $U$ \citep[see][]{Jam09} allows us to get rid of the integral in the last expression and write the joint distribution of $(\bgamma,\bY,\bdelta,U)$, conditionally on $\bX$, as
\begin{align*}
\frac{\alpha^k\,\mathcal{A}_k}{\Gamma(n)} u^{n-1}\e^{-\alpha\psi(u)}\prod_{j=1}^{k} \kappa_{n_j}(u)\,\prod_{j=1}^k G_0(\d \bgamma_j^*)\end{align*}
or equivalently, by using \eqref{eq:short},
\begin{align*}
\frac{\alpha^k}{\Gamma(n)} u^{n-1}\e^{-\alpha\psi(u)}\prod_{j=1}^{k} \kappa_{n_j}(u) G_0(\d \bgamma_j^*)\prod_{i\in C_j} f^*(y_i\mid \bgamma_j^*,\bx_i)^{\delta_i}S^*(y_i\mid \bgamma_j^*,\bx_i)^{1-\delta_i}.
\end{align*}
\end{proof}

\section{Further results from the simulation study of Section \ref{sec:stratdata}}\label{sec:plots}

\begin{figure}[h!]
    \centering
	\subfloat[exact\label{fig:a2}]{\centering\includegraphics[width=0.5\textwidth]{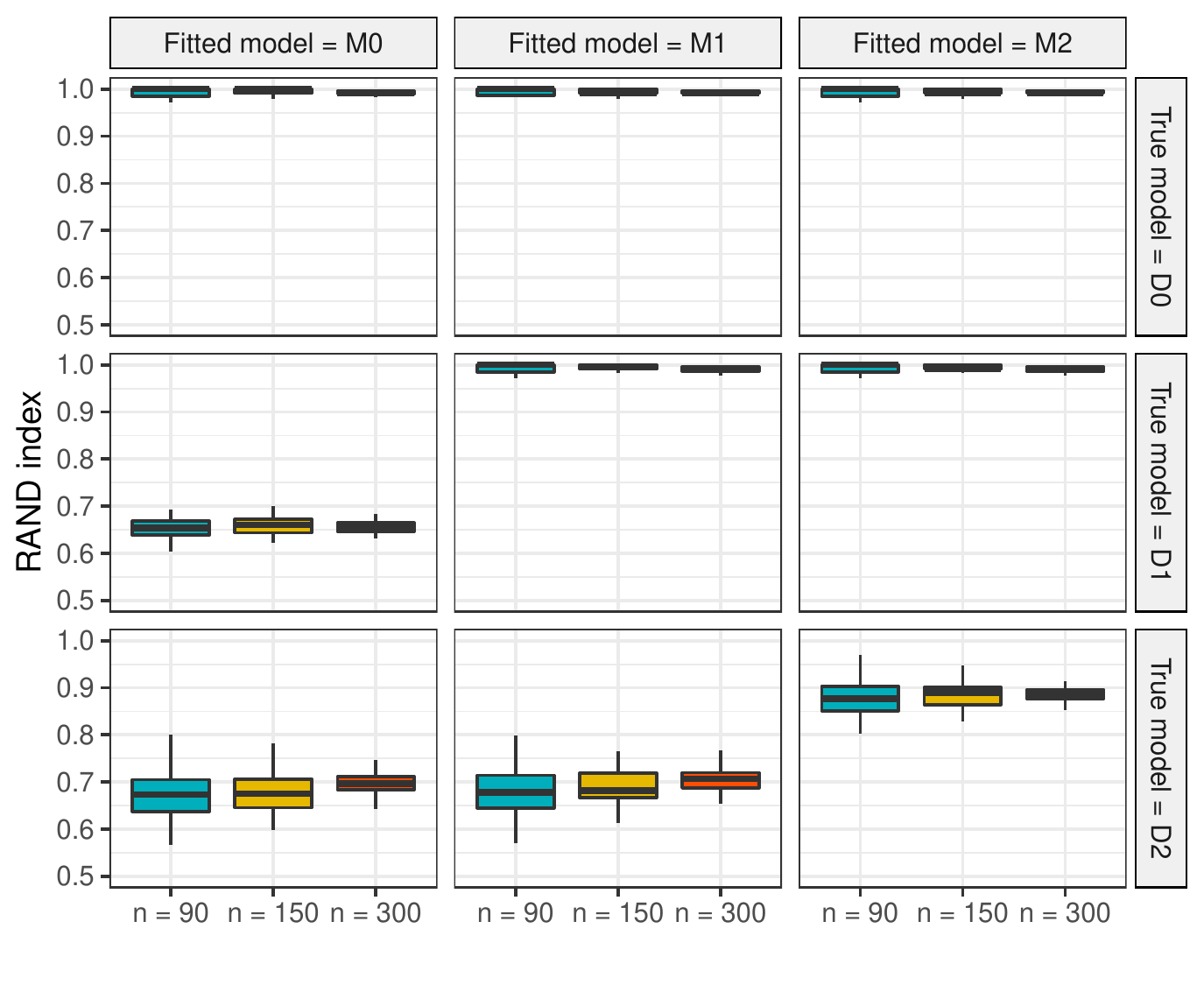}}
	\subfloat[10\% censored \label{fig:b2}]{\includegraphics[width=0.5\textwidth]{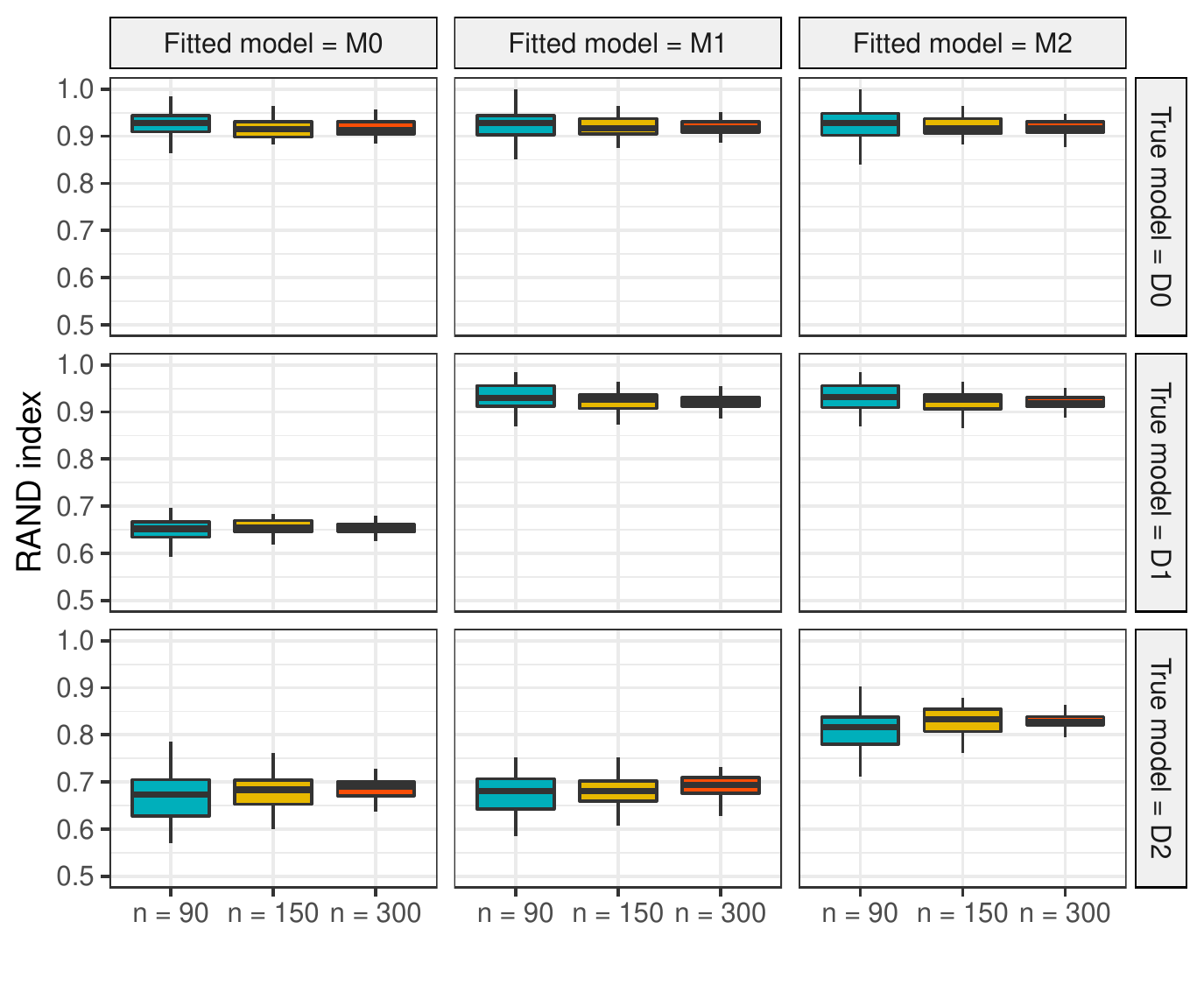}}
	\\[6pt]
	\centering
	\subfloat[20\% censored\label{fig:c2}]{\includegraphics[width=0.5\textwidth]{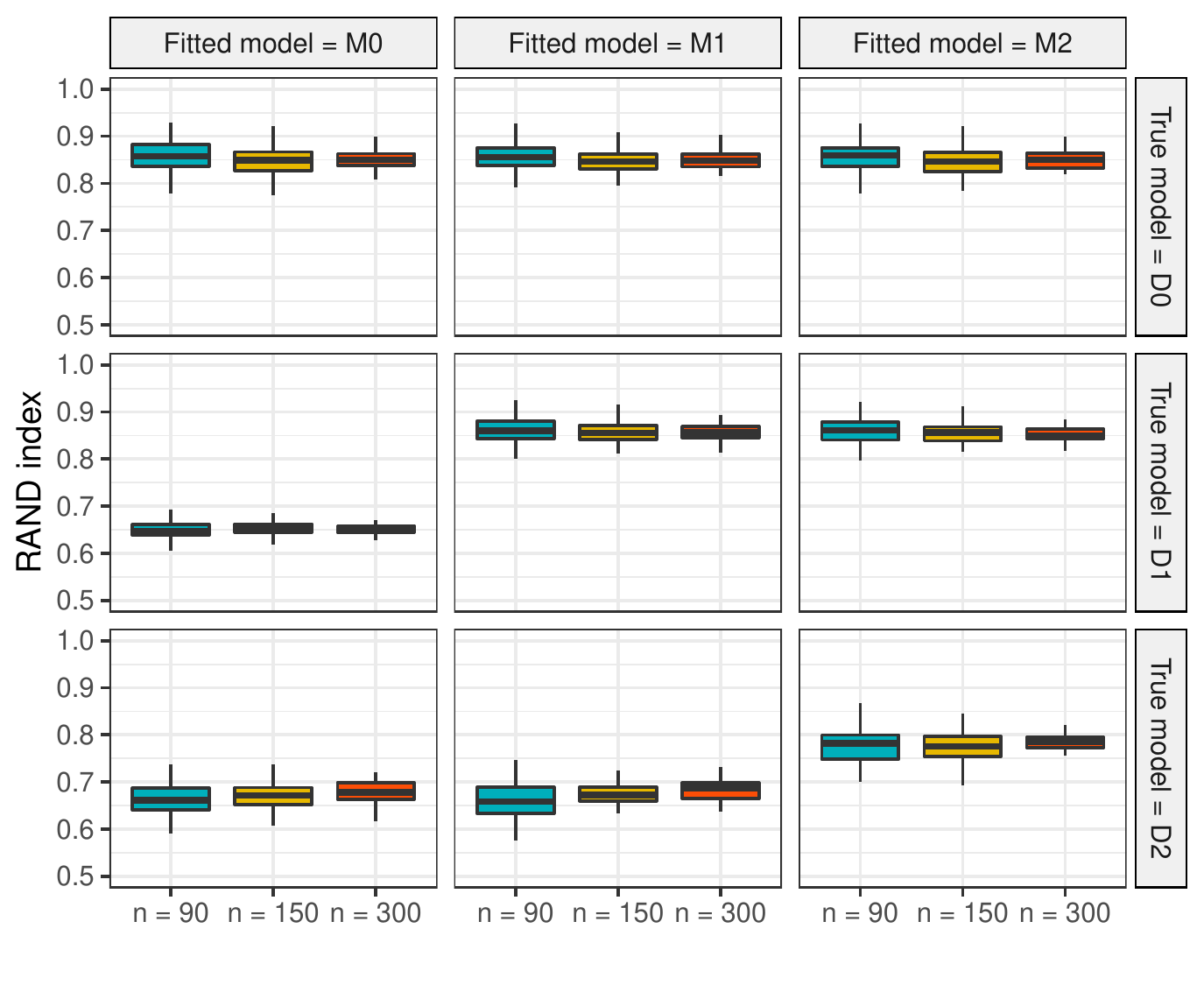}}
	\subfloat[30\% censored\label{fig:d2}]{\includegraphics[width=0.5\textwidth]{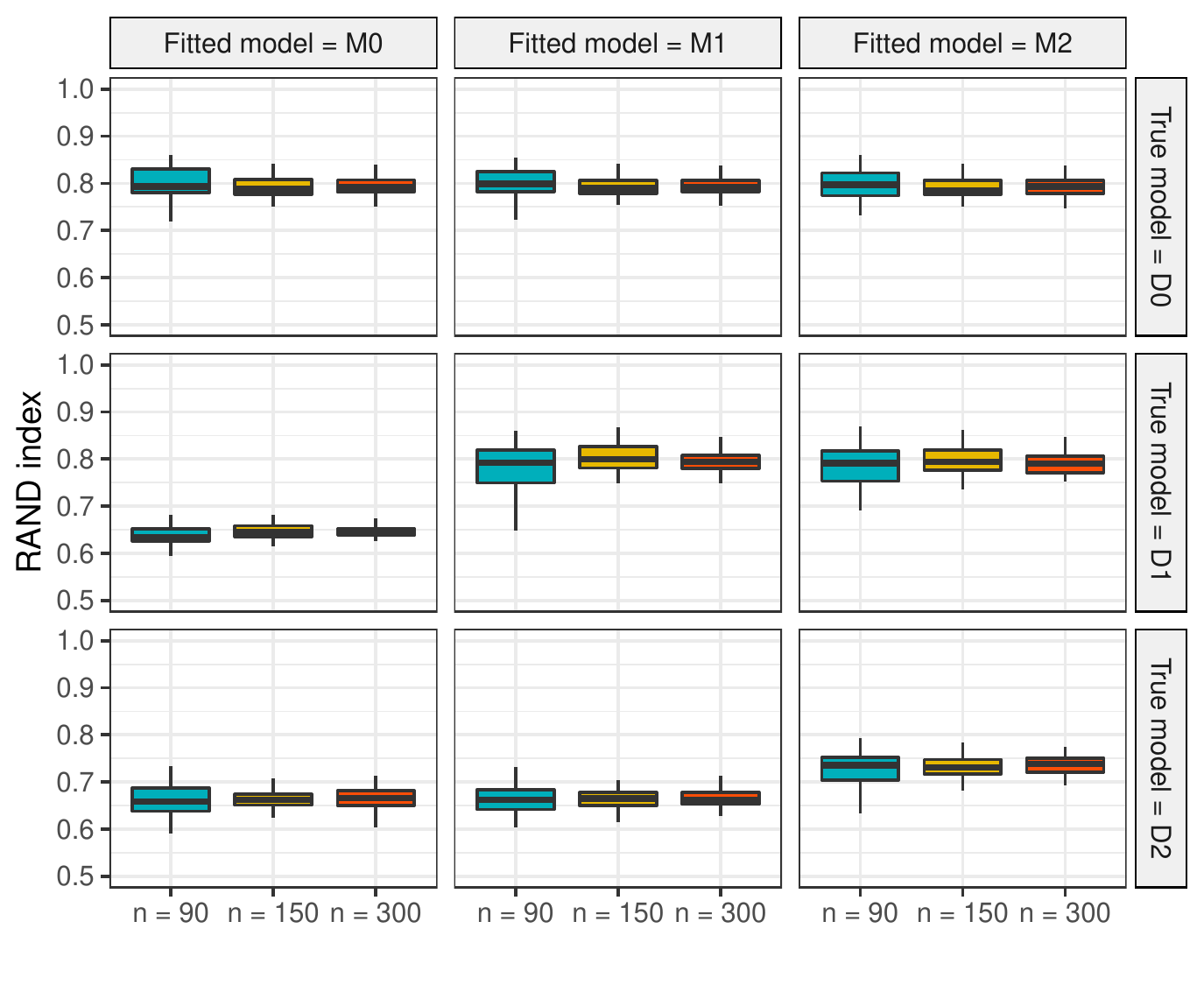}}
\caption{\label{fig:strat_2} Simulated data. RAND index, measuring the similarity between the true partition of the data and the detected optimal stratification, for different sample sizes $n$ ($n=90$ in cyan, $n=150$ in yellow, $n=300$ in red). The boxplots are obtained by analysing 50 replicates of each scenario. The data generating processes are defined by means of a type-I minimum kernel, the models implemented for the log-survival times are defined by means of a logistic kernel. The four panels refer to different percentages of right-censored observations, namely $0\%$ (panel a), $10\%$ (panel b), $20\%$ (panel c) and $30\%$ (panel d). In each panel different rows refer to different data generating processes, different columns refer to different models fitted the data.}
\end{figure}

\begin{figure}[h!]
    \centering
	\subfloat[exact\label{fig:a3}]{\centering\includegraphics[width=0.5\textwidth]{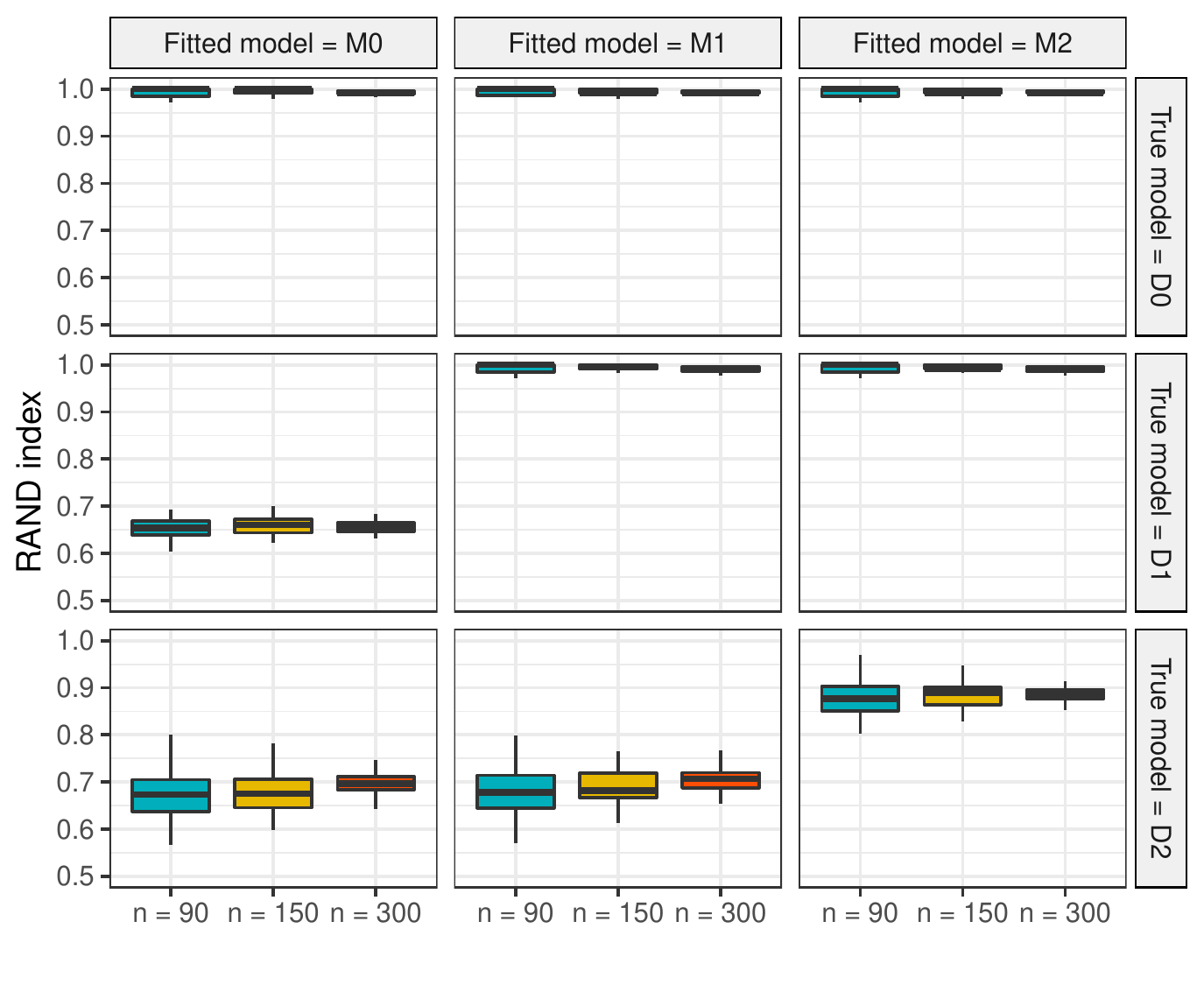}}
	\subfloat[10\% censored \label{fig:b3}]{\includegraphics[width=0.5\textwidth]{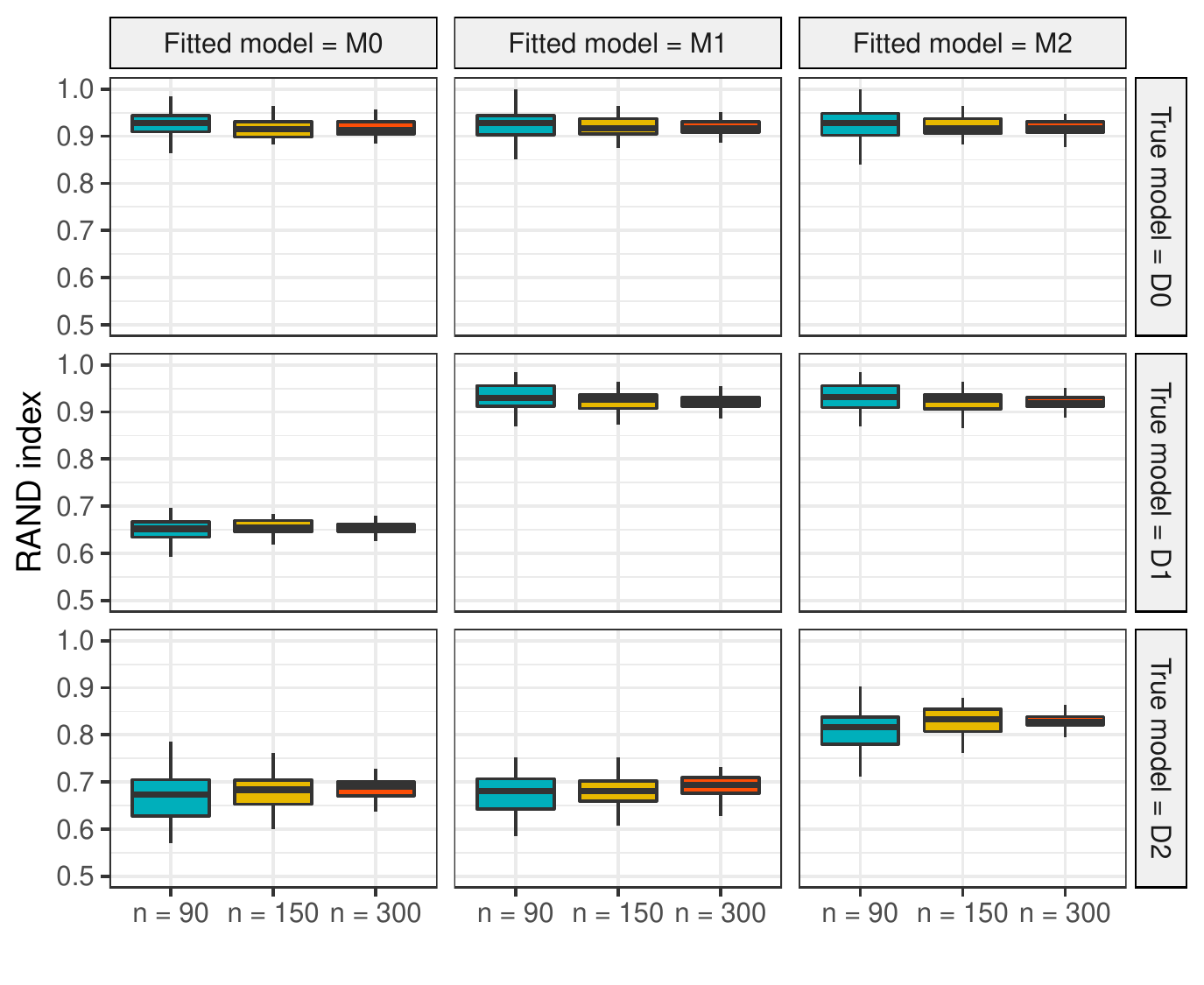}}
	\\[6pt]
	\centering
	\subfloat[20\% censored\label{fig:c3}]{\includegraphics[width=0.5\textwidth]{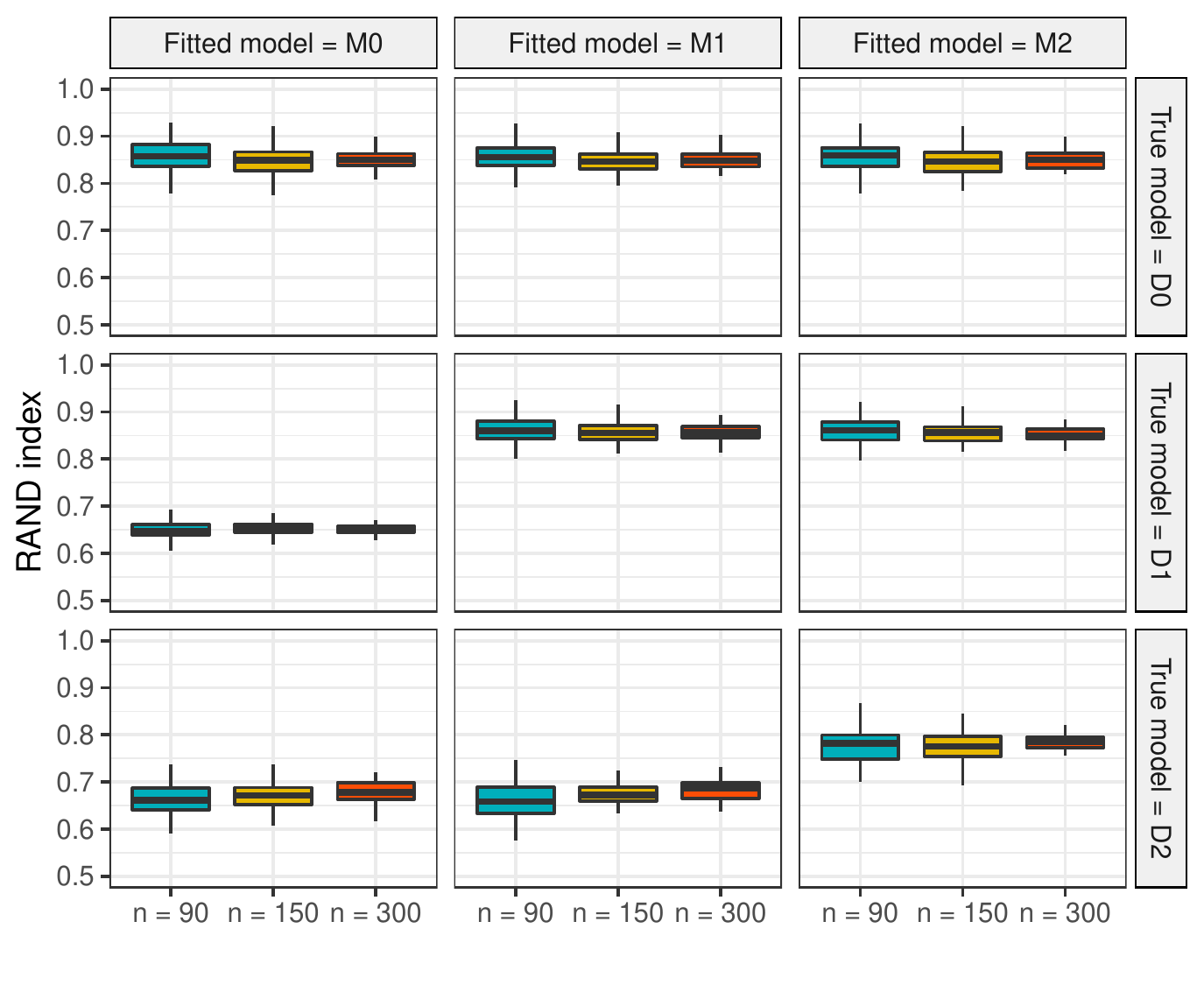}}
	\subfloat[30\% censored\label{fig:d3}]{\includegraphics[width=0.5\textwidth]{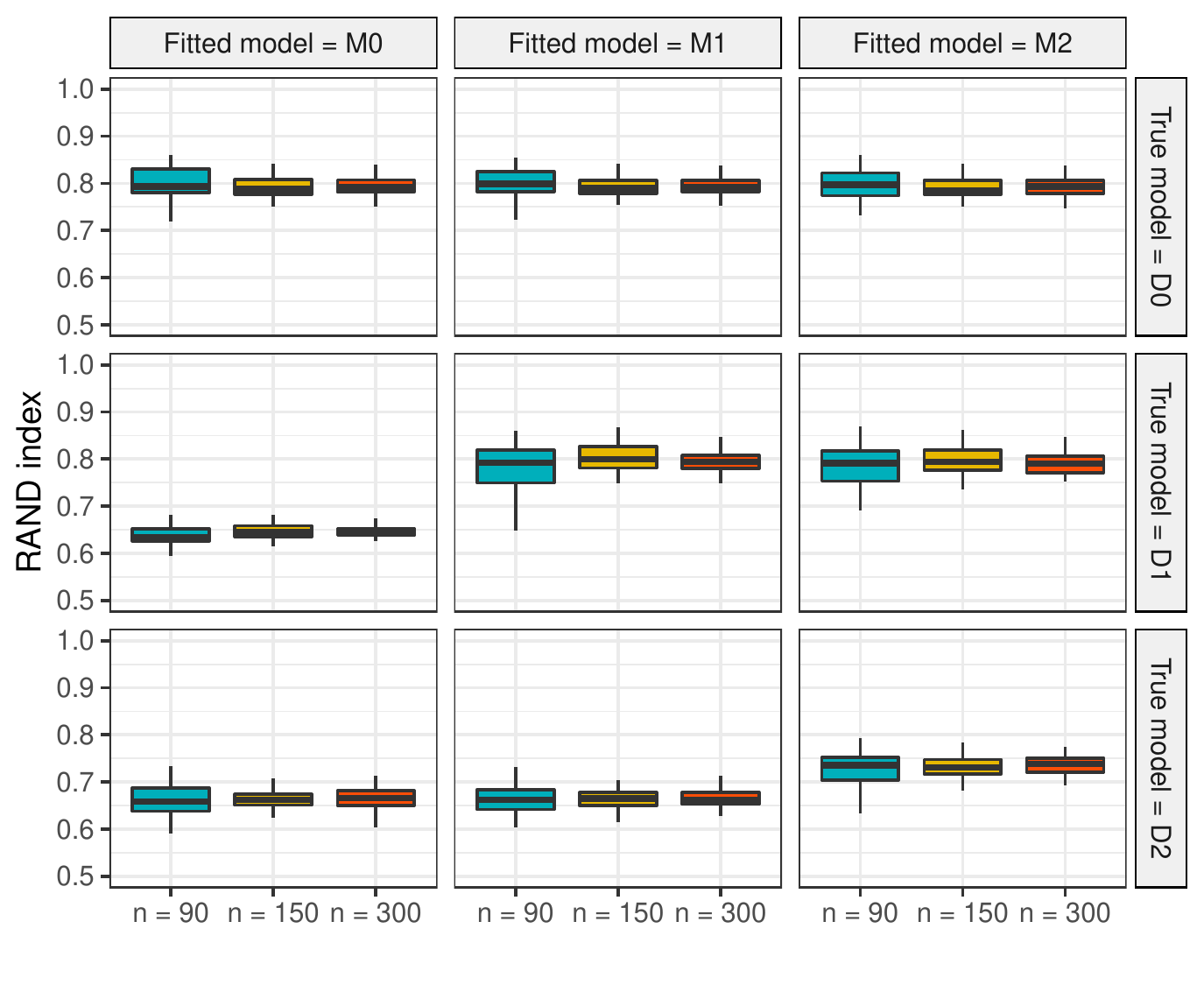}}
\caption{\label{fig:strat_3} Simulated data. RAND index, measuring the similarity between the true partition of the data and the detected optimal stratification, for different sample sizes $n$ ($n=90$ in cyan, $n=150$ in yellow, $n=300$ in red). The boxplots are obtained by analysing 50 replicates of each scenario. The data generating processes are defined by means of a type-I minimum kernel, the models implemented for the log-survival times are defined by means of a normal kernel. The four panels refer to different percentages of right-censored observations, namely $0\%$ (panel a), $10\%$ (panel b), $20\%$ (panel c) and $30\%$ (panel d). In each panel different rows refer to different data generating processes, different columns refer to different models fitted to the data.}
\end{figure}

Figures \ref{fig:strat_2} and \ref{fig:strat_3} display the results of the simulation study of Section \ref{sec:simu}, when data are generated from models with a type-I minimum kernel for the log-survival times, and analyzed with models based on a logistic and normal kernels respectively. 

\begin{figure}[h!]
    \centering
	\subfloat[exact\label{fig:a_strata}]{\centering\includegraphics[width=0.49\textwidth]{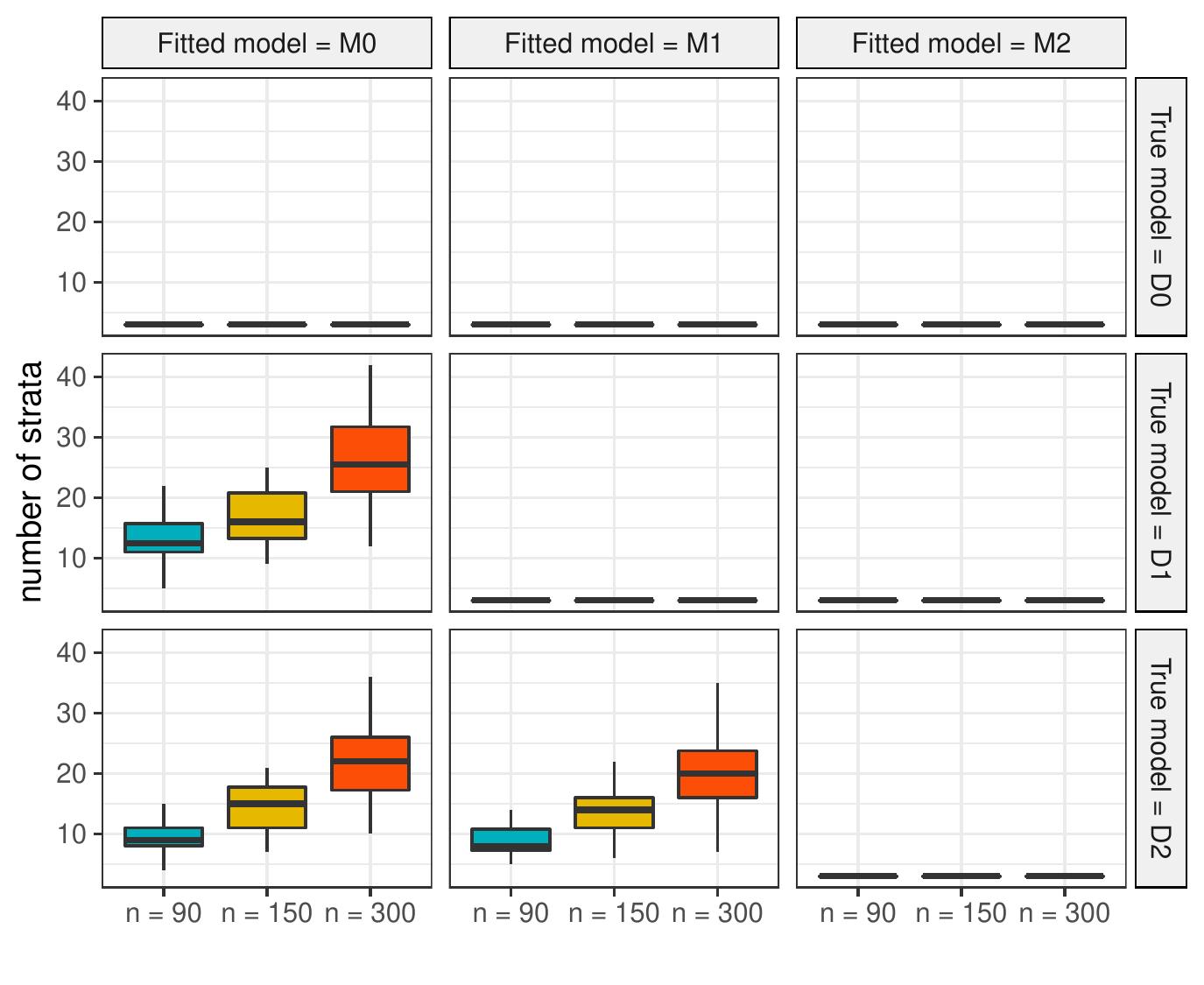}}
	\subfloat[10\% censored \label{fig:b_strata}]{\includegraphics[width=0.49\textwidth]{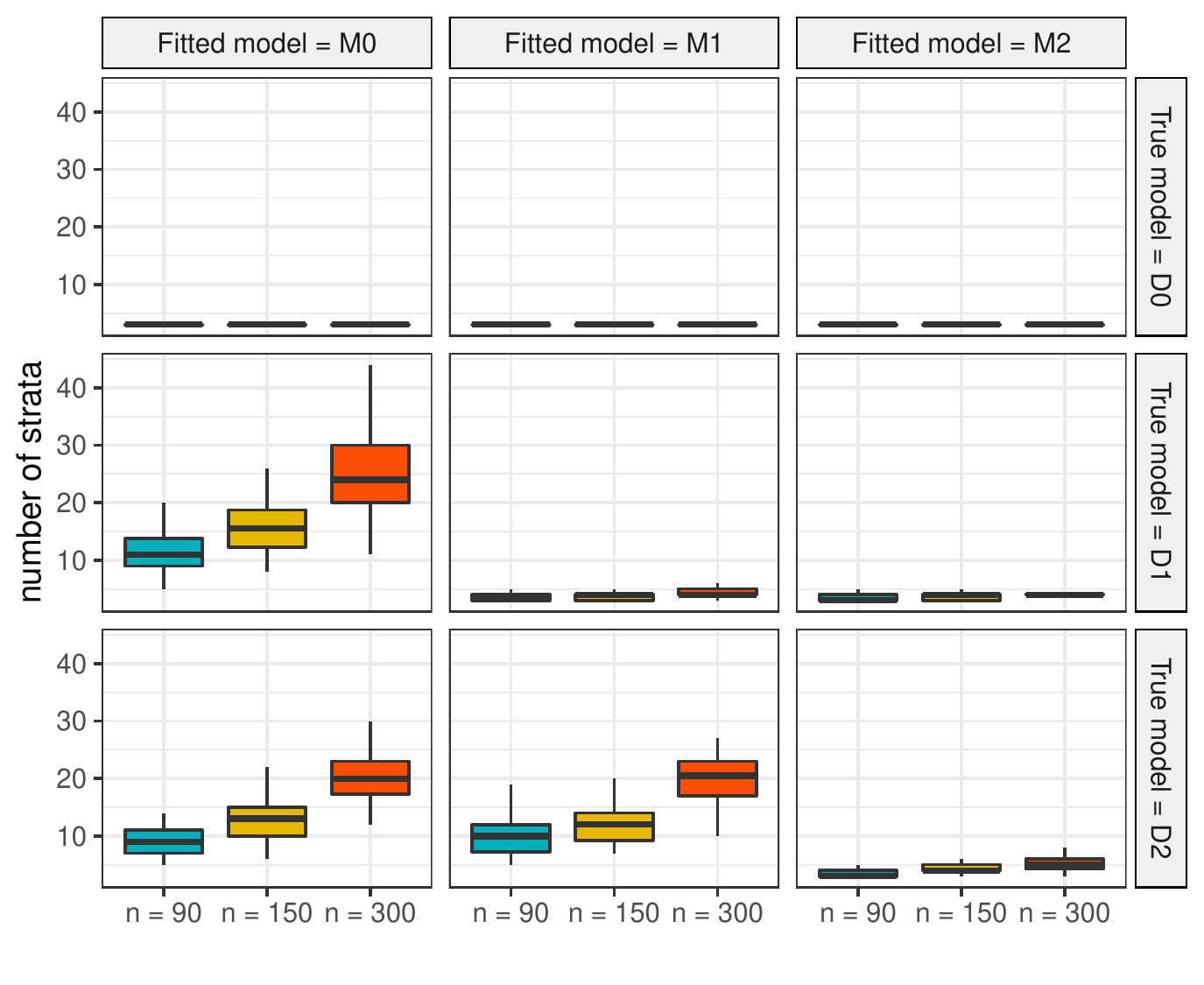}}
	\\[6pt]
	\centering
	\subfloat[20\% censored\label{fig:c_strata}]{\includegraphics[width=0.49\textwidth]{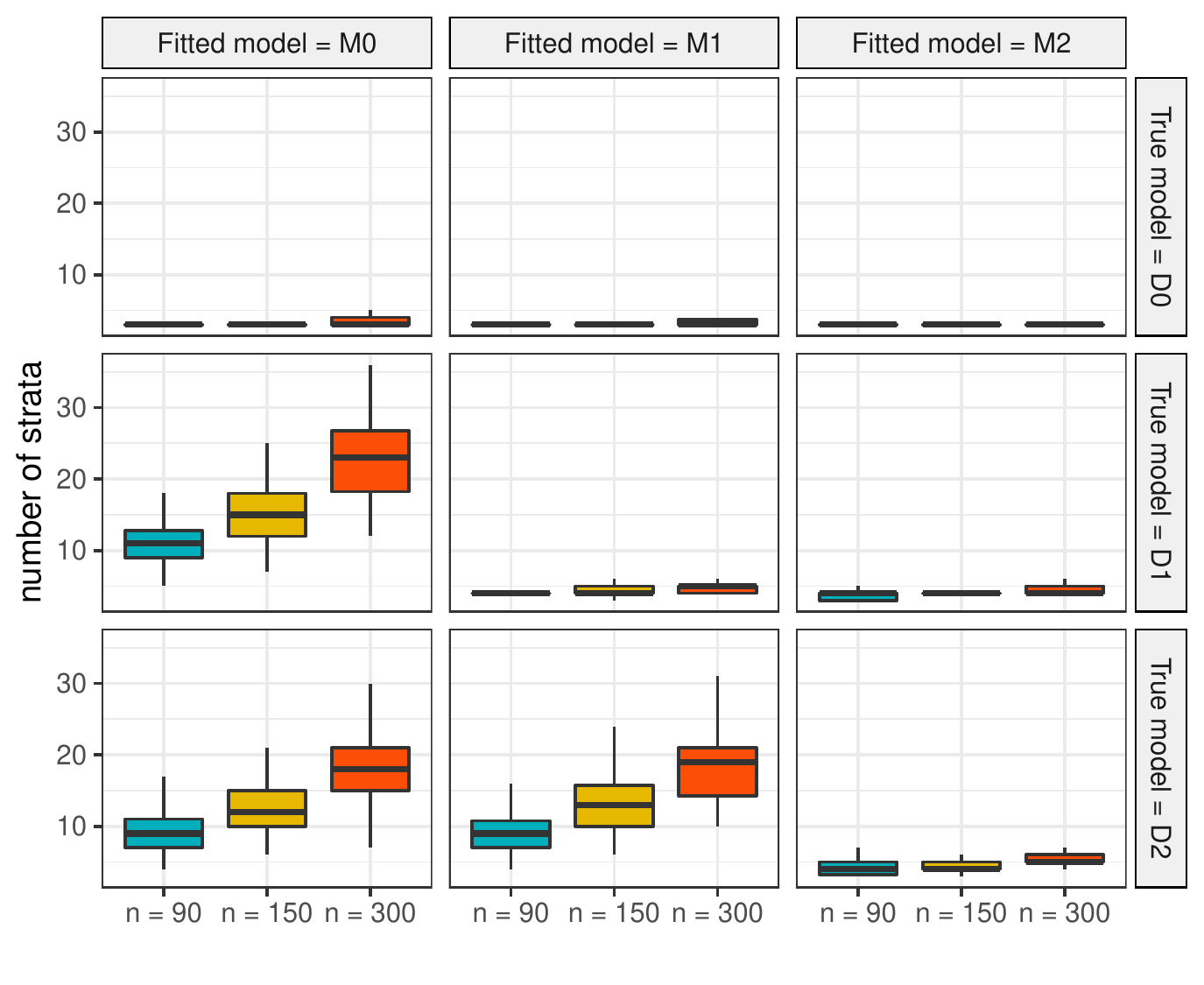}}
	\subfloat[30\% censored\label{fig:d_strata}]{\includegraphics[width=0.49\textwidth]{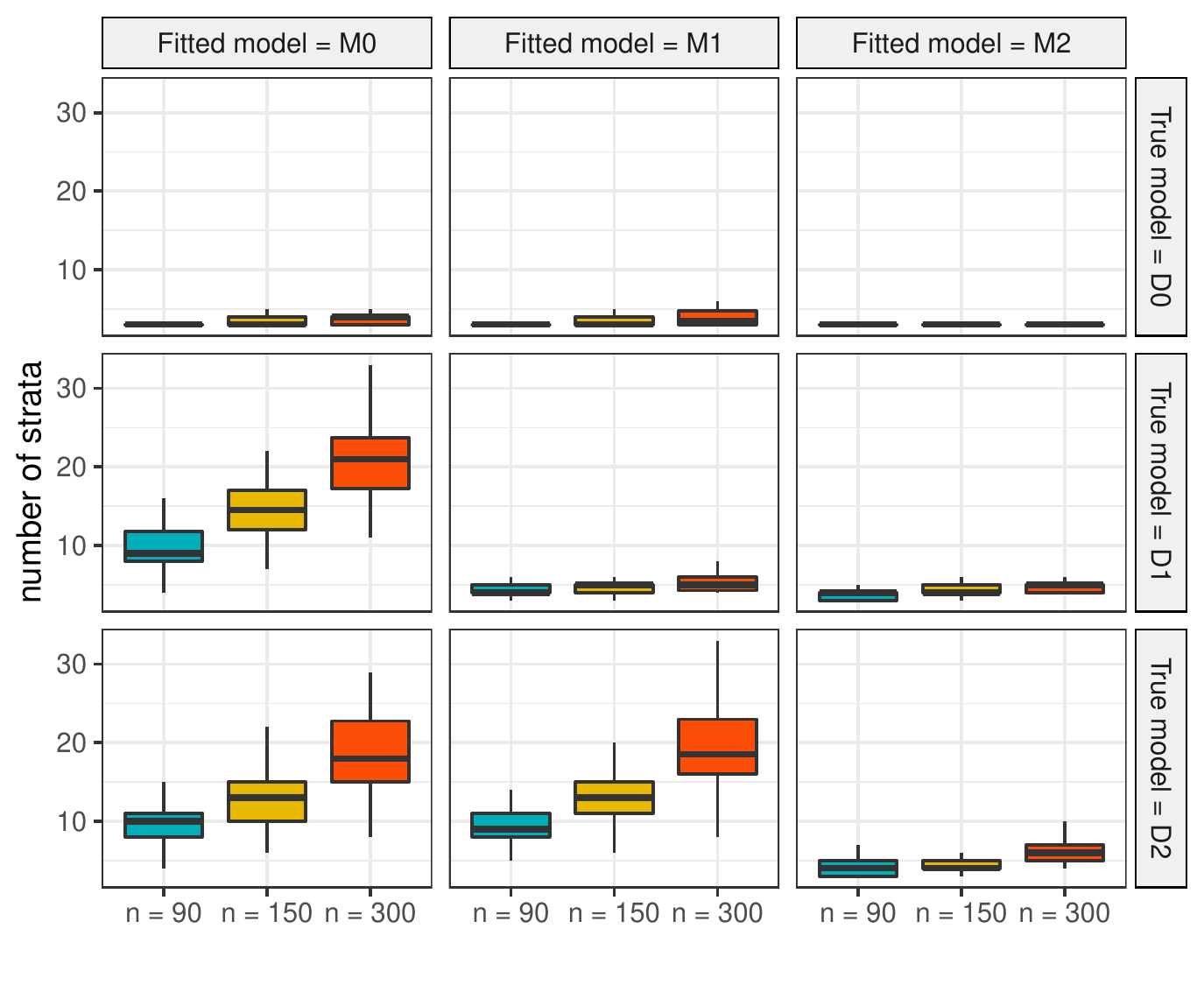}}
\caption{\label{fig:clust_1} Simulated data. Number of identified strata, for different sample sizes $n$ ($n=90$ in cyan, $n=150$ in yellow, $n=300$ in red). The boxplots are obtained by analysing 50 replicates of each scenario. Both the data generating processes and the models for the log-survival times that were implemented, are defined by means of a type-I minimum kernel. The four panels refer to different percentages of right-censored observations, namely $0\%$ (panel a), $10\%$ (panel b), $20\%$ (panel c) and $30\%$ (panel d). In each panel different rows refer to different data generating processes, different columns refer to different models fitted to the data.}
\end{figure}

Figure \ref{fig:clust_1} complements Figure \ref{fig:strat_1} by displaying the number of identified strata for each combination of data generating process D0, D1 or D2, and fitted models M0, M1 or M2, and for different samples sizes $n\in\{90,150,300\}.$

\section{Diagnostics for the analysis of UIS data }\label{sec:app_diag}
\begin{figure}[h!]
    \centering
    \includegraphics[width = 0.3 \textwidth]{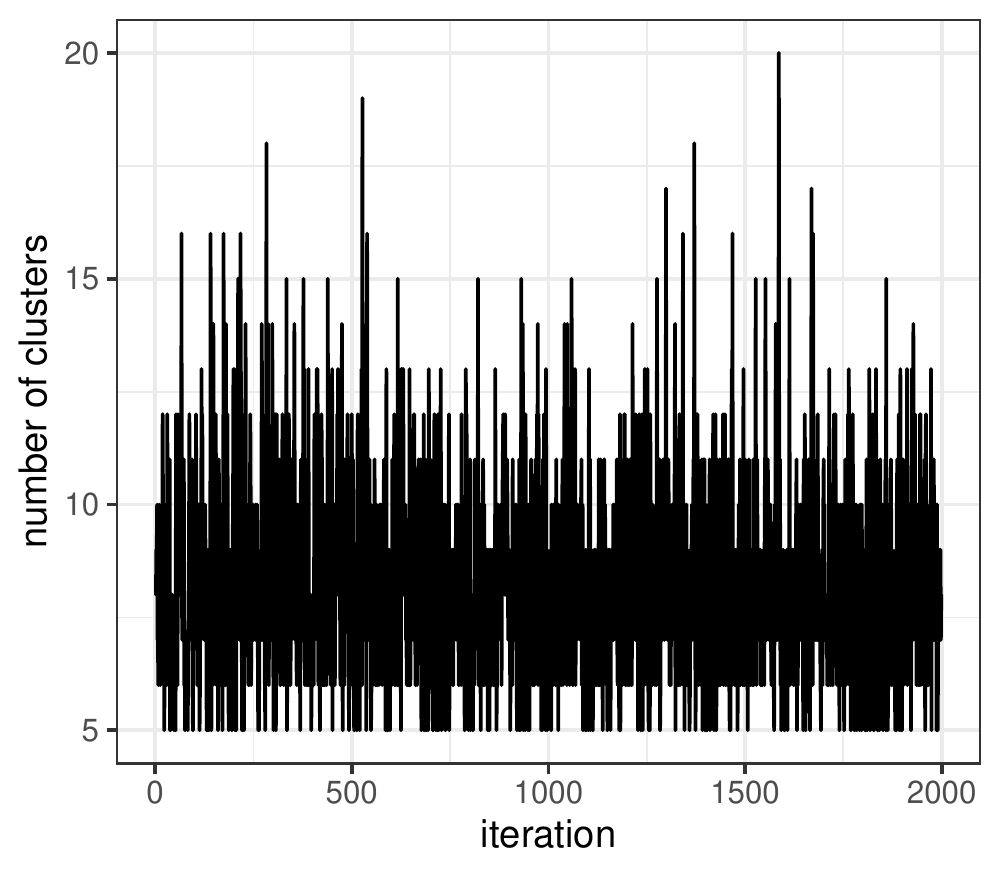}
    \includegraphics[width = 0.3 \textwidth]{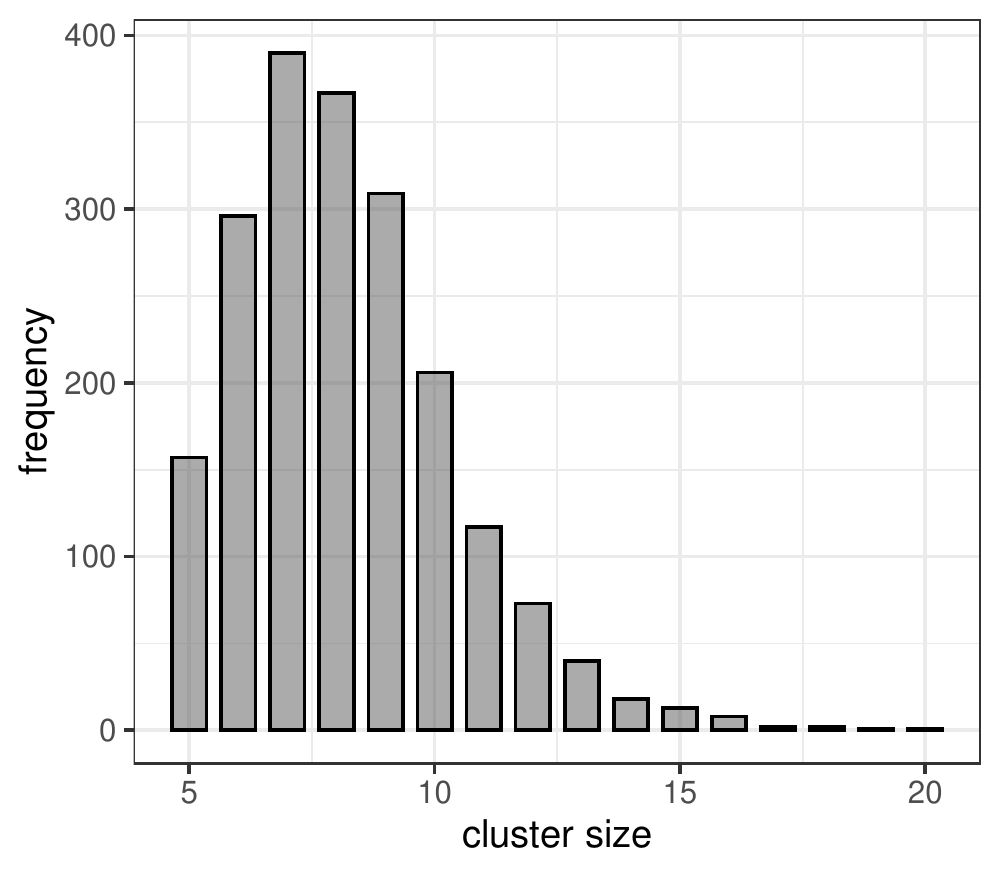}
    \includegraphics[width = 0.3 \textwidth]{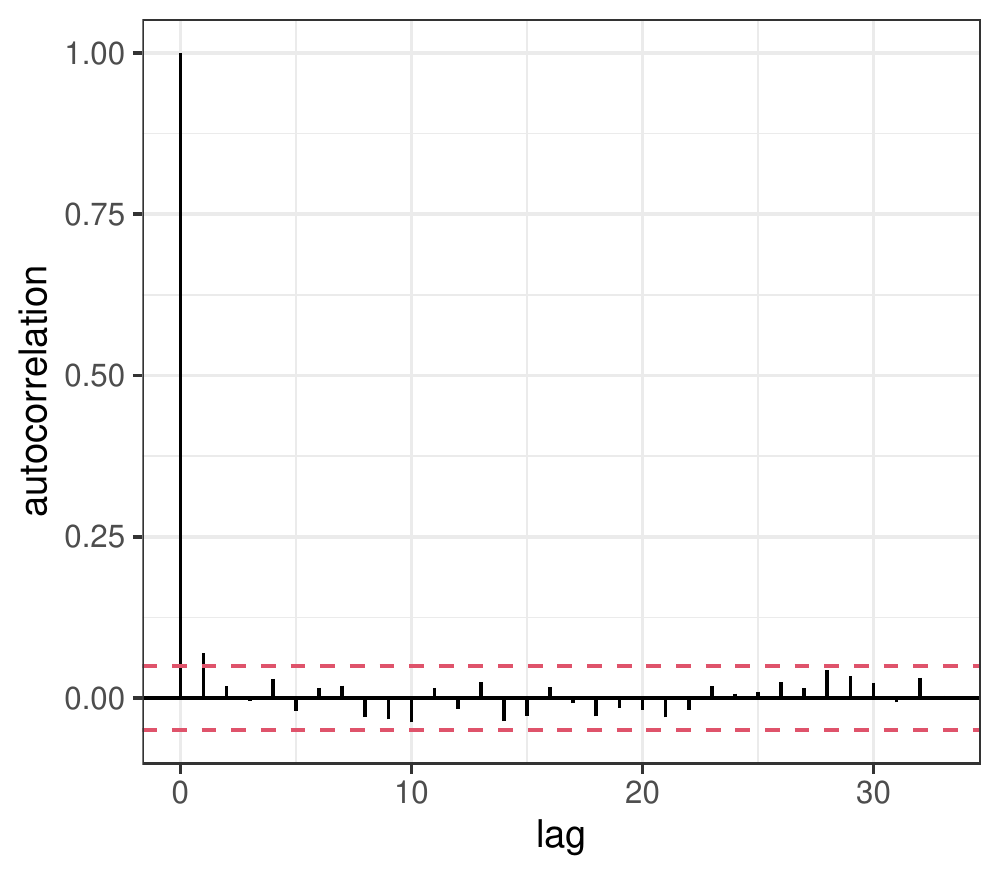}
    \caption{UIS data set. Analysis of number of clusters in the partitions visited by the chain: traceplot (left panel), estimated posterior distribution (middle panel), autocorrelation function (right panel).}\label{fig:diag1}
\end{figure}

We provide some details on the mixing and the convergence of the chain of the algorithm implemented for the analysis of the UIS data set. The acceptance rate for the random walk Metropolis-Hasting steps to update $U$ and $\tau$ are $0.295$ and $0.489$, respectively. Moreover, following a thinning of the chain (keeping one value every ten), the mixing appears satisfactory and the autocorrelation weak. This is displayed in Figure \ref{fig:diag1} which focuses on the number of clusters in the partitions visited by the chain. The right panel of the same plot indicates weak autocorrelation. Finally, the Geweke's convergence diagnostic on the number of clusters is equal to $0.589$, suggesting convergence of the chain to stationarity.

\section{Comparison of N-IG mixtures and DP mixtures for accelerated life models }\label{sec:app_DP}
\begin{figure}[h!]
    \centering
    \includegraphics[width = 0.45 \textwidth]{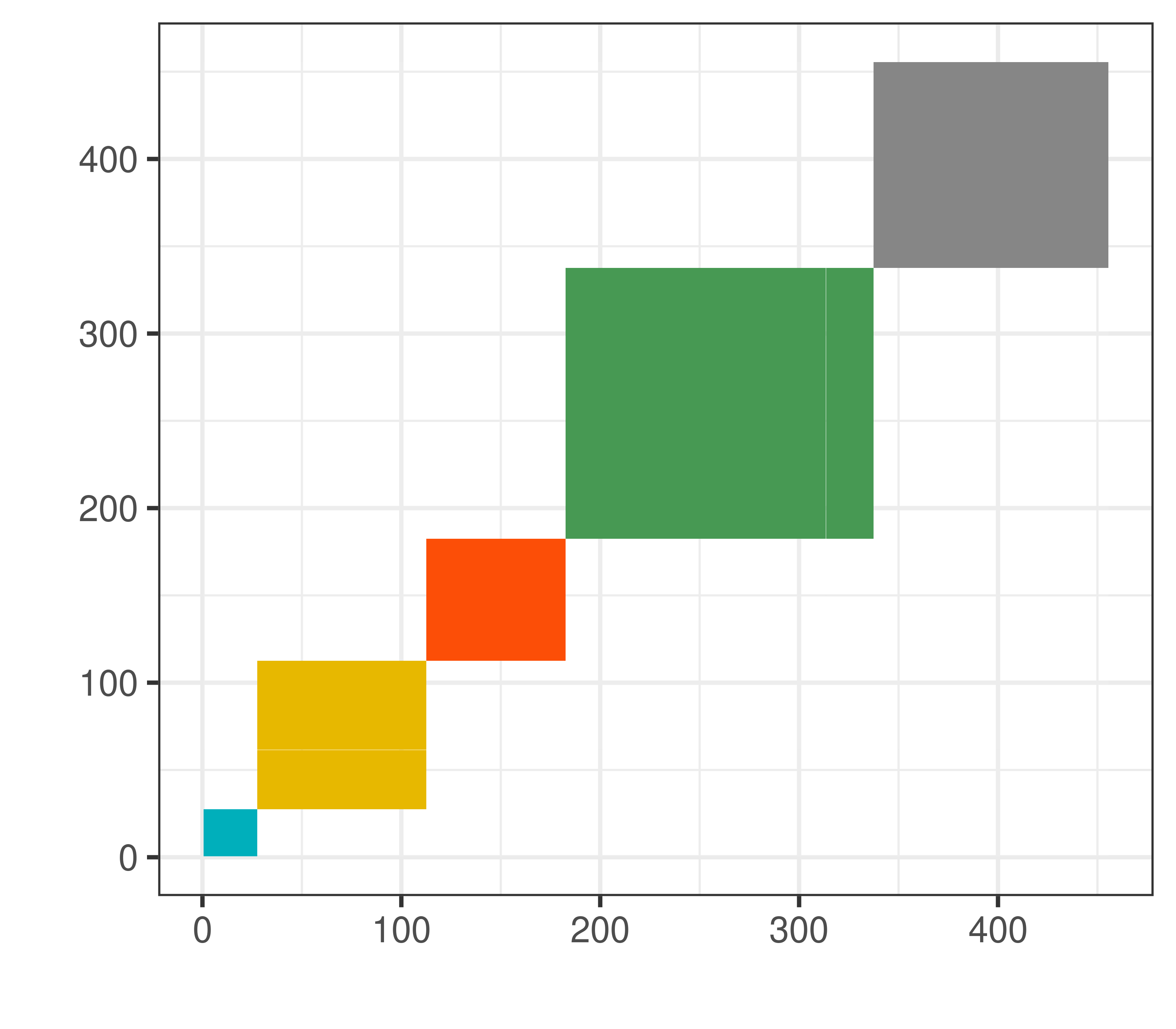}
    \includegraphics[width = 0.45 \textwidth]{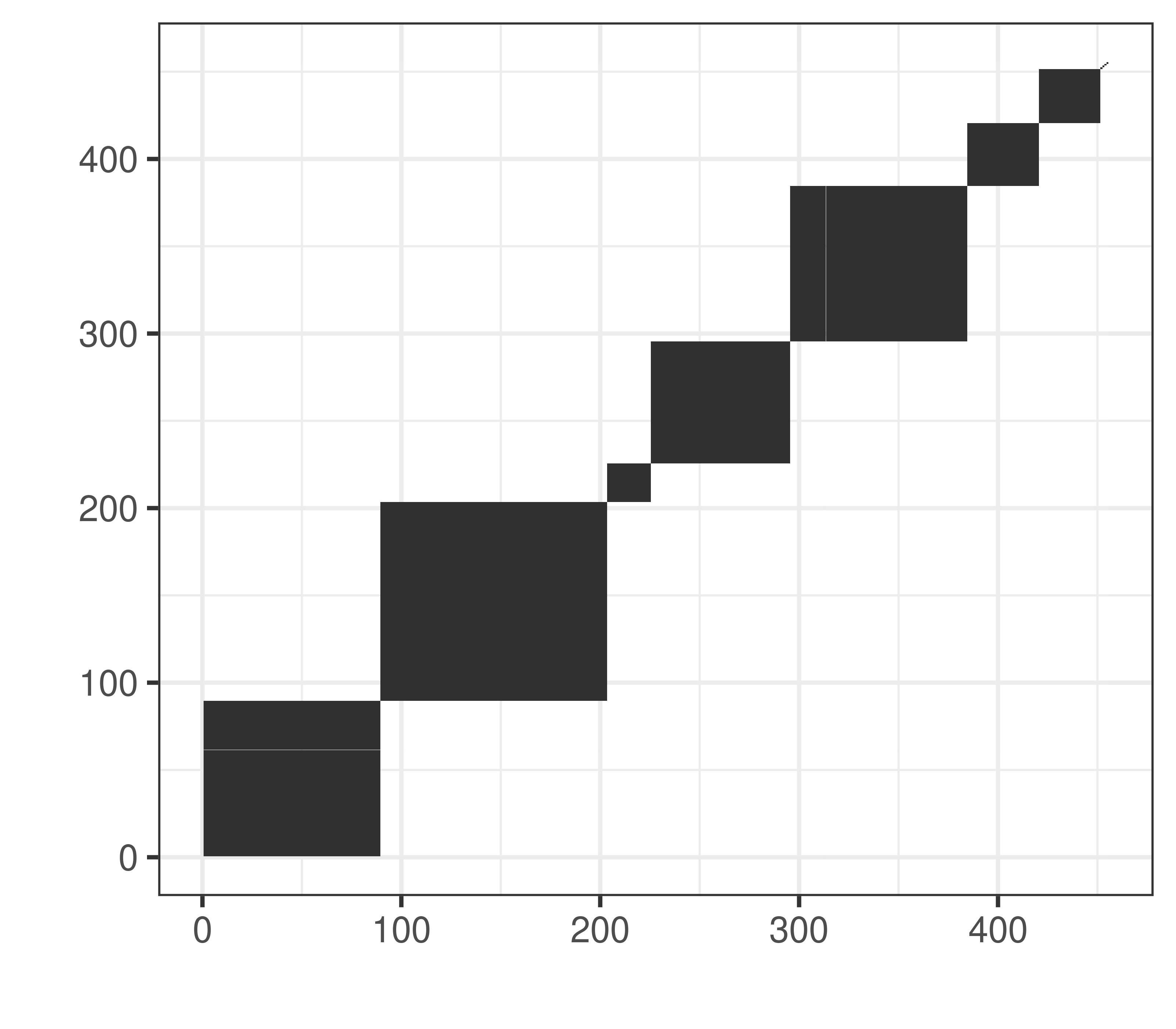}\caption{UIS data set. Left panel: representation of the size of the blocks in the optimal stratification identified with the N-IG mixture model; right panel: representation of the size of the blocks in the optimal stratification identified with the DP mixture model. In both cases, strata are ordered based on the stratum-specific estimated mean expected survival time. The colours of the blocks in the left panel correspond to the ones assigned to each group in Table \ref{tab:frequencies}.}\label{fig:NIG_DP}
\end{figure}

We compared the stratification generated by the nonparametric mixture model with N-IG mixing random probability measure, with the one obtained by completing the same model with a DP mixing measure. In order to make a sensible comparison, we set the total mass parameter of the DP so that the expected value of the prior distribution induced on the number of clusters is the same as the one implied by our specification of the N-IG. The optimal stratification obtained with the DP counts 9 clusters, that is 4 more than the corresponding partition obtained with the N-IG model. The stratification identified via the DP displays two small clusters with frequency respectively equal to 4 and 1. The two plots showed in Figure \ref{fig:NIG_DP} allow us to visually compare the size of the blocks in either identified stratification. 

Overall, the different behaviours displayed by N-IG and DP mixture models seem consistent with the discussion proposed by \citet{Lij05} when comparing N-IG and DP. 

\section{Pitman--Yor case}\label{sec:PY}
We believe it is worth studying the performance of other classes of mixture models, not considered in this work, in the context of possibly censored survival data. A model which does not belong to the framework we set forth, but which has gained considerable popularity in applications \citep[see, e.g.,][]{Cor20,carmona&al:19} thanks to its flexibility and tractability, is the Pitman--Yor mixture model \citep{per:92}. 
The same strategy presented in our work can be devised by completing models \eqref{eq:bnpmm0}, \eqref{eq:bnpmm1} and \eqref{eq:bnpmm2}, with the assumption that $G$ is distributed as a Pitman--Yor process. To this end, we present a result, analogous to Proposition \ref{prop:posterior}, which represents the starting point to derive the full conditional distributions needed to implement a Gibbs sampler for the case of Pitman--Yor mixture models.

\begin{proposition}
	\label{prop:posteriorPY}
	Let $(Y_i,\delta_i,\bX_i)$, $i=1,\ldots,n$, be a set of observable random variables from model M2. Let $G$ be distributed as a Pitman--Yor process, $G\sim PY(\theta, \sigma, P_0)$, such that $\sigma \in [0,1)$ and $\theta>-\sigma$. Then the joint conditional density of 
	$(\bgamma,\bY,\bdelta)$ conditionally on $\bX$ is given by
	\begin{align}
	\nonumber
	\frac{\prod_{j=1}^{k-1}(\theta+ j\sigma)}{(\theta+ 1)_{n-1}}\prod_{j=1}^k(1 - \sigma)_{n_j -1}\prod_{i\in C_j} \{f^*(y_i\mid \bgamma_j^*,\bx_i)\}^{\delta_i}\{S^*(y_i\mid \bgamma_j^*,\bx_i)\}^{1-\delta_i},
	\end{align}
	where the sets $C_j=\left\{i\in\{1,\ldots,n\} : \bgamma_i =\bgamma_j^*\right\}$.
	
\end{proposition}

Proposition~\ref{prop:posteriorPY} is proved by combining the steps of the proof of Proposition~\ref{prop:posterior} with the construction of the Pitman--Yor process in terms of completely random measures given by \citet{Pit97}. 

\end{document}